\documentclass[isre,nonblindrev]{informs3} 
\OneAndAHalfSpacedXI 

\usepackage{natbib}
 \bibpunct[, ]{(}{)}{,}{a}{}{,}%
 \def\bibfont{\small}%
\TheoremsNumberedThrough     

\EquationsNumberedThrough    

\usepackage{geometry}
\usepackage{graphicx}
\usepackage{float,graphicx}
\usepackage[graphicx]{realboxes}
\usepackage{caption}
\usepackage{booktabs}
\usepackage{float}
\usepackage{amsmath}
\usepackage{subcaption}
\usepackage{booktabs}
\usepackage{adjustbox}
\usepackage{amssymb} 
\usepackage{tabularx}  
\usepackage{multirow}
\usepackage{booktabs}

\usepackage{xcolor}
\usepackage{subcaption}
\usepackage{makecell}

\begin{document}

\TITLE{Platform Design, Earnings Transparency and Minimum Wage Policies: 
Evidence from A Natural Experiment on Lyft\footnote{This study was conducted as an independent academic research project. No consulting fees, research grants or other payments have been made by Lyft to the authors. We thank Lyft for sharing a fascinating data set with us. We also thank Yannis Bakos, Jeremy Bird, Chris Conlon, Sarah Cormack-Patton, Zoe Cullen, Peter Day, Lawrence De Geest, Andrew Garin, Jerry Golden, Liz Hahn, David Holtz, John Horton, Saahil Karpe, Ramayya Krishnan, Siddharth Patil, Stefano Puntoni, David Risher, Maher Said, Chris Sholley, and seminar participants at Lyft, the 2024 Workshop on Information Systems and Economics, the 2025 Marketing Science Conference, the 2025 BU Platform Symposium, the 2025 INFORMS Conference on Information Systems and Technology, Boston University, Carnegie-Mellon University, Hong Kong University of Science and Technology, Nanyang Technological University, the National University of Singapore, Purdue University, the University of Illinois at Urbana-Champaign, the University of Notre Dame, and the University of Texas at Austin for helpful discussions and feedback on preliminary versions of these results. Financial support from the Fubon Center for Technology, Business and Innovation and from the NET institute (www.NETinst.org) is gratefully acknowledged. 
} 
}
\ARTICLEAUTHORS{
  \AUTHOR{Rubing Li, Xiao Liu, Arun Sundararajan} 
  \AFF{New York University, Leonard N. Stern School of Business}
}

\ABSTRACT{
We study the effects of a significant design and policy change at a major ridesharing platform that altered both provider earnings and platform transparency, examining how it altered outcomes for drivers, riders and the parent platform, and providing managerial insights about how to balance competing stakeholder interests while avoiding unintended consequences. In February 2024, Lyft introduced a policy guaranteeing drivers a minimum fraction of rider payments while also increasing per-ride earnings transparency. The rollout of this policy was staggered, first introduced in “major markets” that were more urban, providing a natural experiment to assess how platform transparency and earnings guarantees affect ridesharing availability, driver engagement and rider satisfaction. Using trip-level data from over 47 million rides in a major urban market and its neighboring suburban markets across six months, we applied dynamic staggered difference-in-differences models along with a geographic border strategy to measure the causal effects of these platform design changes on supply- and demand-side outcomes, ride production and platform outcomes. We show that the design change led to substantial changes in driver engagement, with separate effects from the guarantee and the transparency. Drivers increased their working hours and utilization, leading to more completed trips and higher per-hour and per-trip earnings. These effects were strongest for drivers with lower pre-policy earnings and greater income uncertainty. We unpack the economic mechanism by which these changes led to a positive spillover on demand. We also provide some evidence that points to platform transparency potentially leading to unintended strategic driver behavior. In ongoing work, we outline a counterfactual simulation framework that models ride production as a function of driver supply hours and rider intents, examining how small changes in driver choices might have amplified the positive effects of the intervention even further, and develop a self-supervised machine learning model that leverages driver trajectory embeddings to predict multihoming behavior and examine whether supply increases came from both the expansion of driver activity as well as substitution from competing platforms. Our study shows how platform-led interventions present an intriguing alternative to government-led minimum pay regulation, and provide new strategic insights into managing platform change. 
}
\maketitle
\newpage

\section{Introduction}

Over the last decade, platform business models have created new ways of organizing economic activity across industries ranging from transportation and car rental to short-term accommodation and local retailing (\cite{parker2017platform}; \cite{sundararajan2017sharing}). As platforms like Airbnb, Doordash, Getaround, Lyft, Turo and Uber have grown and evolved, they have defined a new relationship form with their providers that use their platform, one that is clearly distinct from the traditional employment relationship that characterized 20th century interactions between companies and workers, and which is shaped by a range of platform design decisions (\cite{filippas2023limits}). Examples of such decisions may include the choice of pricing levels, whether the platform or its providers that use their platform set these prices, how providers and passengers are matched,  how platforms and providers share revenues, and the extent to which providers are aware of this division.

There have been many platform design changes related to pricing on app-based ride-hailing platforms like Uber and Lyft over the last decade  A key innovation early in the evolution of ridesharing was the adoption of  innovative dynamic algorithmic pricing mechanisms, tying price levels to supply and demand. Although revenue sharing arrangements between platforms and providers that use their platform were initially fairly simple — the provider would receive a fixed percentage of the price paid by a passenger — over time, pricing and revenue division became more complex as variety of taxes, benefits surcharges, congestion fees and other collective dues (like the NY State Black Car Fund contributions) were mandated by different jurisdictions. In parallel, US cities like New York enacted regulations that set mandatory minimum pay rates which used complex formulas that aggregated per-minute and per-mile tariffs to dictate a floor on driver earnings. While ride-hailing platforms still continued to choose pricing levels, these regulatory minimums significantly constrained the extent to which the market mechanism could shape pricing. Put differently, in such regulated markets, while market forces dictate the prices that consumers were willing to pay for ride-hailing in what remains a highly competitive market, driver earnings are constrained in parallel by regulatory minimums. 

A natural alternative to minimum pay laws is to let market forces dictate both consumer pricing and the division of revenues. Indeed, there are a wide variety of other platform design choices — including how consumers search, how providers and passengers are matched, and what determines provider entry and exit, that platforms continue to determine, and which promote the efficiency and smooth functioning of this new economic form.

In this paper, we study the effects of a platform design change that altered both minimum earnings and per-ride transparency into the division of revenue between a platform and providers that use their platform.  Our work provides insights into the evolution and eventual outcomes associated with a major platform design change, and examines the effects of this kind of platform-led alternative to government-led minimum pay regulation.

In February 2024, Lyft (henceforth referred to as “the platform”) announced a new earning standard, under which drivers would be guaranteed to earn 70\% or more of rider payments each week, after external fees (in what follows, we refer to this as the “earning guarantee”), and under which the platform would provide greater trip-level transparency about what the rider paid, and the breakdown of the amounts that were paid to the driver, the platform, and to other government entities or collectives (in what follows, we refer to this as “earning transparency”).  This earning guarantee and earning transparency feature was rolled out to drivers in a selected set of major markets in February 2024, and then, three months later, was rolled out nationwide.  This move by the platform towards an earning guarantee and earning transparency represented a major shift in platform policy that potentially altered driver preferences for utilizing the platform relative to other earning opportunities. In what follows, we refer to this platform design change as the “intervention.”

Ex ante, it was not clear how this intervention would be received by drivers. While a greater level of transparency fosters trust and gives drivers superior information they can learn from and subsequently better optimize their platform activity it could also have countervailing effects by inducing negative reactions from those drivers because of variation in the trip-by-trip revenue split with the platform. Furthermore, while some drivers could perceive the weekly earnings share “floor” of 70\% as generous, others who had historically earned a higher weekly fraction might have perceived the move as an attempt by the platform to increase their take rate. 

Design decisions and changes like what the platform undertook are central strategic choices for two-sided platforms, and especially for two-sided transaction platforms for which indirect network effects are particularly pronounced. A flawed design decision can lead to either customer or provider backlash and reduce participation on one side of the platform, causing ripple effects on the other side and rapidly leading to a negative spiral. Indeed, there is a history of unanticipated consequences to significant platform changes. For example, when TaskRabbit changed their pricing mechanism from an auction system to one involving fixed hourly rates, there was significant backlash from the providers, many of whom left the platform \footnote{https://www.theverge.com/2014/6/17/5816254/taskrabbit-blows-up-its-auction-house-to-offer-services-on-demand}. When a popular peer-to-peer car rental platform shifted from provider-led to platform-led pricing in a manner that significantly increased provider revenue, this change nevertheless induced an unexpected negative reaction from providers and led to increased levels of churn (\cite{filippas2023limits}).

The positive or negative ripple effects of such design choices can be especially rapid for platforms like Uber, Lyft and Doordash that offer real-time two-sided transactions. The strategic importance of such design choices, the uncertainty around their economic effects, and the intriguing platform-led alternative this specific change offered to government-mandated minimum earnings policies motivated our study. The staggered rollout of the earning guarantee and earning transparency created a natural experiment that we use to understand how platform policy shapes provider behaviors.  We analyze how this move to an earning guarantee and the provision of earning transparency changed different aspects of the behavior of drivers on the platform, the overall demand patterns for the platform, and the ensuing ride production. We also analyze how this shift varied by driver and market characteristics, the extent to which changes altered platform and rider outcomes, and explored whether supply changes might have reflected new driver time or substitution from other platforms. 

We conduct our analysis on a unique proprietary dataset obtained from the platform. This dataset comprises anonymized detailed trip-level data comprising about 47 million rides for one large major market — LAX — that was part of the initial 2024 three-month rollout, and from a set of two adjacent non-major markets — OCX and SBD —  that were not, for the 25-week period surrounding the February 2024 launch, and for the same window of time in 2022-23. This data is combined with market-level data about app opens, ride requests and conversion rates, as well as with selected historical data about drivers.

We use a series of models to analyze the nature of changes in driver and market outcomes caused by this policy shift. We begin with a  baseline set of dynamic staggered difference-in-differences models (\cite{callaway2021difference}).  In these baseline models, we balance the treated and control cohorts of drivers using inverse propensity weighting, and restrict the sample to those drivers whose pre-treatment driving patterns were sufficiently inclusive of the major (LAX) market. We also estimate alternative models, including one that use a geographic border strategy to better account for self-selection, a model that uses 02/23-05/23 outcomes as the control on a subset of drivers whose pre-treatment (11/23-02/24) driving behaviors were sufficiently unchanged relative to the same period the previous year (11/22-02/23), and a simpler two-way fixed effects model focusing on just the first week of the intervention.  Based on a ride production function we have estimated, we also examine whether small changes to driver starting location and time might have improved the impact of the intervention, and based on a new machine-learning based model that predicts whether a driver is multihoming on other platforms, we examine the difference in behavior changes between those drivers who provide exclusively for Lyft and those who drive for other platforms as well. 

Across our models, we  find  that the effect of the earning guarantee and earning transparency on driver outcomes was positive. 

\begin{itemize}
    \item The new platform features led to a significant increase in driver engagement. We estimate that the intervention caused an increase in the weekly hours drivers spent driving for Lyft by between 8.6\% and 33.3\%.
    \item Treated drivers also completed more trips per week and drove more sessions per week, with each session also lasting longer, and more trips being completed per hour. 
    \item Treated drivers also earned more per hour and per trip. 
\end{itemize}

These effects were more pronounced for drivers that belonged to specific groups
\begin{itemize}
    \item Drivers whose “take rate” was lower than median prior to being treated, and who had lower uncertainty tolerance displayed higher treatment effects. This difference is consistent with the earning guarantee and earning transparency interventions having different effects on driver behaviors, and consistent with the idea that the variance reduction effect of the weekly earnings guarantee was received more favorably by selected drivers than the transparency effect.  
    \item Full-time drivers displayed higher treatment effects than part-time drivers. For example, the increase in working hours was 26\% greater for full-time drivers, compared to the increase for part-time drivers. 
\end{itemize}

Our preliminary results (not reported in detail in this paper) also suggest that drivers who multihome displayed higher treatment effects even after controlling for platform activity levels, consistent with a household production theory that the supply changes are more readily substituted from a competing platform rather than from leisure. 

We establish that the supply shock spilled over into increasing demand, or put differently, establish a causal effect of the changes in driver activity on increasing the rate at which riders open their app and consider booking a ride, and the rate at which riders convert requested rides into completed rides. This establishes that there was a \textit{market expansion} effect that added to the positive effects of higher utilization rates. We explore how the fact that  these demand effects are concentrated in a small number of extremely “high demand” areas could lead to countervailing effects of the intervention on different metrics of inequality. 

We also briefly discuss ongoing work that unpacks the “growing of the pie” by examining whether the increase in supply and separately, the increase in demand, led to not just an increase in the production of rides, but an increase in the \textit{efficiency} with which rides were produced. We do so by estimating a ride production function that takes supply (driver hours) and demand (app opens and ride requests) as inputs. Our preliminary results show that ride production displays increasing returns to scale, and furthermore, that the efficiency of ride production increased due to the intervention.  Since driver supply cannot simply be “reallocated” in a counterfactual experiment because the supply in a location depends in part on where prior riders wanted to be dropped off, we have also developed a novel counterfactual strategy that computes the expected changes in supply that results from shifting driver \textit{starting locations}, and we are using it to explore whether the increases in ride production could have been further improved through simple behavioral nudges that shifted driver starting locations and times by small amounts. Our preliminary results suggest that small \textit{location} shifts lead to greater returns than small \textit{time} shifts. 

Overall, our research and findings validate a bold platform decision, and establish that driver benefits did not come at the cost of lower platform earnings or inferior rider experiences. Rather, the intervention grew the pie in multiple ways and led to a win-win-win outcome across platform stakeholders, suggesting that platform-led earning floors may serve as a viable substitute for externally imposed minimum wage policies, an approach better aligned with a data-driven delegation approach to regulation (\cite{Cohen2015SelfRegulationAI}). This seems especially important in a dynamic setting where market forces dictate the prices that consumers are willing to pay, which makes it more likely that if driver earnings are constrained by regulatory minimums, productive efficiency and social surplus will be compromised. 

We organize the rest of this paper as follows. Section 2 discusses related literature and the connection of our findings to prior work. Section 3 describes the platform design change in a little more detail, our data and how we aggregate and organize it. Section 4 describes our empirical strategy and how we use a variety of models to address specific issues of endogeneity and self-selection.  Section 5 presents our results, Section 6 describes ongoing work, and we conclude and outline future work in Section 7. 

\section{Related Literature}

Like many other two-sided transaction platforms, ride-sharing platforms leverage a range of technology that significantly improves market efficiency by optimizing matching processes and resource allocation (\cite{einav2016p2p}). For example, dynamic pricing leads to greater capacity utilization (\cite{cramer2016disruptive, hall2017uber}), improve service quality through reduced wait times (\cite{hall2017uber, athey2019service}), and increase consumer surplus (\cite{cohen2016bigdata}). On the supply side, ride-sharing platforms enable flexible labor markets (\cite{einav2016p2p}), allowing drivers to choose when, where and how much they work (\cite{hall2017uber, chen2019value}), and reducing matching frictions (\cite{castillo2024surge}), while also enabling other positive spillovers that reduce driving while inebriated (\cite{greenwood2017ridesharing}) and traffic congestion (\cite{erhardt2019congestion}).

As the key lever that shapes provider and customer choices, the platform’s choice of pricing mechanism and revenue sharing is of central strategic importance. This is especially important given the fact that the structure of two-sided network effects (\cite{parker2005twosided}) for ridesharing platforms is complex and localized, that indirect network effects can be instrumental in determining competitor entry and inter-platform dynamics (\cite{zhu2012entry}), and small changes in pricing and revenue sharing that alter supply dynamics can lead to equilibrium changes that indirect network effects amplify. As a regulated industry not subject to the economics of network effects, the taxi industry relied on fixed medallion leases and government-mandated fare structures. In contrast, ride-sharing platforms set prices dynamically and implement revenue sharing models that involve collecting a commission that is a fixed or variable fraction of what drivers earn after successful transactions. Such a structure reduces entry barriers for drivers and enables greater flexibility (\cite{chen2019value, angrist2021uber}), while also mitigating potential moral hazard by aligning driver incentives with the platform’s goals (\cite{liu2021moralhazard}). Sophisticated data science-driven mechanisms like dynamic pricing—and in particular, surge pricing—enable real-time adjustments to fares based on fluctuating demand, optimizing utilization during peak hours. The welfare effects of surge pricing have been studied both theoretically and empirically (\cite{cohen2016bigdata, cachon2017surge, castillo2023matching, castillo2024surge}) and the mechanism has been used to understand the nature of shifts in market equilibrium (\cite{hall2017uber, Hall_Horton_Knoepfle_2023}).  A broader literature on platform design choices has investigated how different aspects of the choice of a pricing mechanism can shape platform outcomes. For example, \cite{filippas2023limits} examined the welfare effects of centralized, decentralized and hybrid division of pricing responsibility between providers and the platform in the context of peer-to-peer car sharing, demonstrating that hybrid control can often dominate centralized pricing when one takes provider satisfaction and churn into account.\cite{brown2023competition} documented how pricing frequency, commitment and asymmetry in price-setting technology influence competitive dynamics among firms.  

The policy change that we focus on is closely tied to the concept of transparency in economic and organizational settings. A vast literature in economics identifies several dimensions of transparency that play critical roles in shaping agent behavior and competitive dynamics; some studies that focus on platforms and compensation are particularly relevant to our context. For example, studies on operational transparency—the visibility of platform operations or processes, such as how decisions are made or services are matched—have shown that enhances user perceptions of effort and trust (\cite{buell2011labor, buell2016service}) across various settings, including ridesharing and crowdfunding (\cite{mejia2019crowdfunding, mejiaTransparencyFails, buell2016service}). While we do examine transparency in earnings, our work is distinct from the literature on  transparency in compensation, which focuses on salary disclosure within organizations and the ensuing perceptions of inter-agent (rather than principal-agent) fairness perceptions and associated implications for employee behavior such as wage bargaining (\cite{cullen2024pay, cullen2023equilibrium}). Our study is also related to algorithmic transparency, revealing the decision-making process made by algorithmic systems, which addresses critical issues of accountability, fairness, and biases (\cite{datta2016algorithmic}) while introducing new dynamics that influence stakeholder behavior and platform operations (\cite{wang2023algorithmic}). Finally, in our ongoing work, we also examine how the policy may have unintended consequences on inequality, as it disproportionately benefits certain types of drivers and certain geographic regions over others. This analysis connects to the broader literature on how digital marketplaces and uneven access to digital information can amplify disparities and contribute to platform-induced inequality (\cite{brynjolfsson2006longtail, sen2020socialdistancing, li2024telehealth}).

\section{Data and Context}

\subsection{Context}

In what follows, we refer to the platform design change as the intervention. Each week, every driver on the platform receives a summary of their earnings on the platform. Prior to the rollout of the intervention, drivers would receive a breakdown in this weekly summary of the net passenger payments generated by their rides, along with the driver’s earnings. This information was not provided to the drivers on a ride-by-ride basis — rather, for each ride, a driver would simply be informed about their earnings from the ride, and would not be given any information about the price the passenger paid for the ride, or the net passenger payments from the ride. 

\begin{figure}[ht]
    \centering
    \includegraphics[width=0.9\linewidth]{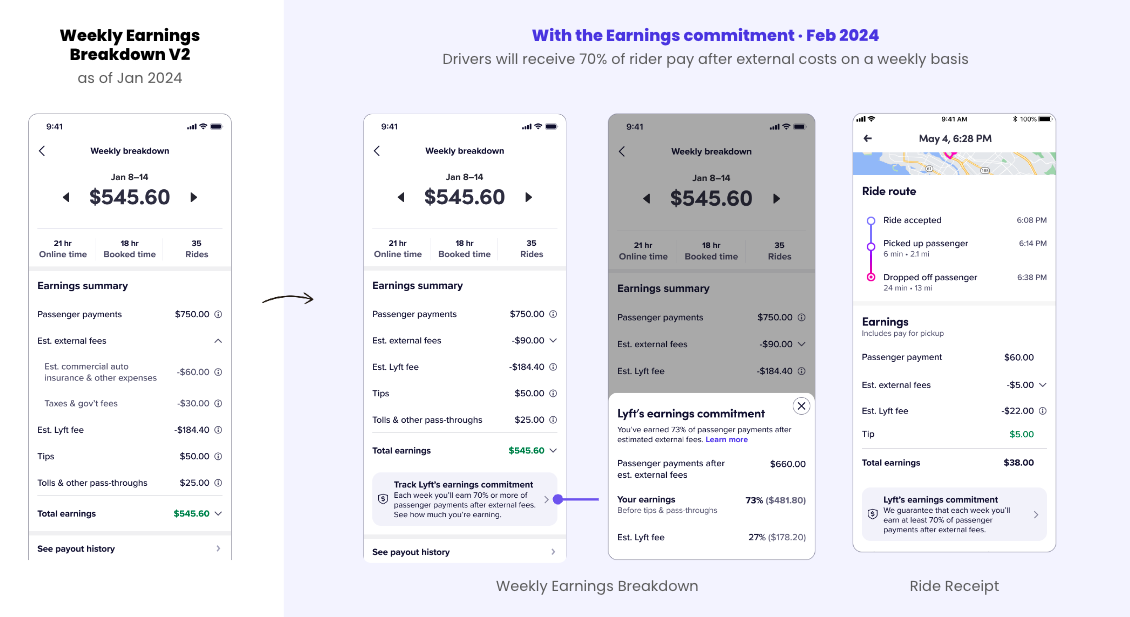}
    \caption{\centering Changes to trip-level transparency and information about the earnings commitment on the driver app after a driver enters the treatment group. }
    \label{fig:interface}
    \vspace{-2em}
\end{figure}

The intervention enacted the following changes(also see Figure \ref{fig:interface}):
\begin{enumerate}
    \item The platform guaranteed that the driver would receive at least 70\% of the net passenger payments generated from the rides they provided in any given week. If a driver is under 70\% of net passenger payments at the end of the week, they are paid an additional amount equal to  the difference between their earnings and the guaranteed minimum amount. The net passenger payment is calculated by subtracting all government fees, benefit fees and other mandatory charges passed through to external parties from the total payments received from passengers who took rides with the driver. A driver’s weekly earnings report highlighted the earning’s commitment and the exact percentage received by the driver, with a new feature that the driver could click on to see a calculated percentage of the net passenger payments from their rides that they were paid. See panels 1 and 2 of the earnings commitment screenshots in Figure  \ref{fig:interface}. 
    \item For \textit{each ride},  the driver was provided with information about the total payment from the passenger, the external fees, the platform’s take, and the net earnings. This represented a significant transparency change; prior to the rollout of the intervention, a driver would not have visibility into what the rider paid, and would only be informed about the driver’s own earnings from the trip. See panel 3 of the earnings commitment screenshots in Figure \ref{fig:interface}.

\end{enumerate}

Change (1) corresponds to what we call the \textit{earning guarantee}, and change (2) to what we call \textit{earning transparency}.  The platform partitions the land area of the continental US into a set of non-overlapping “markets.” Let’s call the set of markets that the intervention was initially rolled out in the major markets and the set of markets that the intervention was not rolled out in the adjacent markets. Any driver who ended a trip in a major market was automatically and immediately added to the set of drivers who received the earning guarantee and earning transparency, and this guarantee and transparency was then applied to all their subsequent trips across all markets they drove in. 

Thus, active drivers who drove primarily or exclusively in major markets were “treated” with the earning transparency and the earning guarantee almost immediately after the intervention was launched. A driver was notified about the changes to their earnings as soon as they made this first dropoff. For simplicity, we also refer to this as being added to the “treatment group.” Additionally, there were some drivers who drove primarily in adjacent markets but who happened to end a trip in a major market sometime during the initial rollout window. These drivers were added to the treatment group at different points in time during the initial rollout window. Finally, there were some drivers who drove exclusively in adjacent markets during the initial rollout window and thus remained in what we term the “never-treated” group.

\subsection{Data and Variables}

We obtained data about one large major market (LAX), and two large adjacent markets (OCX and SBD) that are not characterized as major markets. We refer to these as our target markets. (Further details about the exact markets and the basis for their choice are available in Appendix A.) For these target markets, we analyze data for the time periods November 14th, 2023 to May 6th, 2024, and November 14th 2022 to May 6th, 2023. In each of these time windows, we analyze data about every single trip accepted by every driver that either originated or ended within one of our target markets. The data include a unique driver id,  timestamps of trip request/start/end time, vehicle type, certain geographic information about origin and destination, ride distance, wait time, driver earnings, platform commission, tip amount and rating information. The geographic information is sufficiently granular for us to ascertain when a driver entered the treatment group. This data is combined with geography-specific (zip code level) indicators of overall demand on an hourly basis and price levels on a weekly basis in all the target markets, which we describe in more detail later in the paper. 

\begin{figure}[ht]
    \centering
    \includegraphics[width=0.8\linewidth]{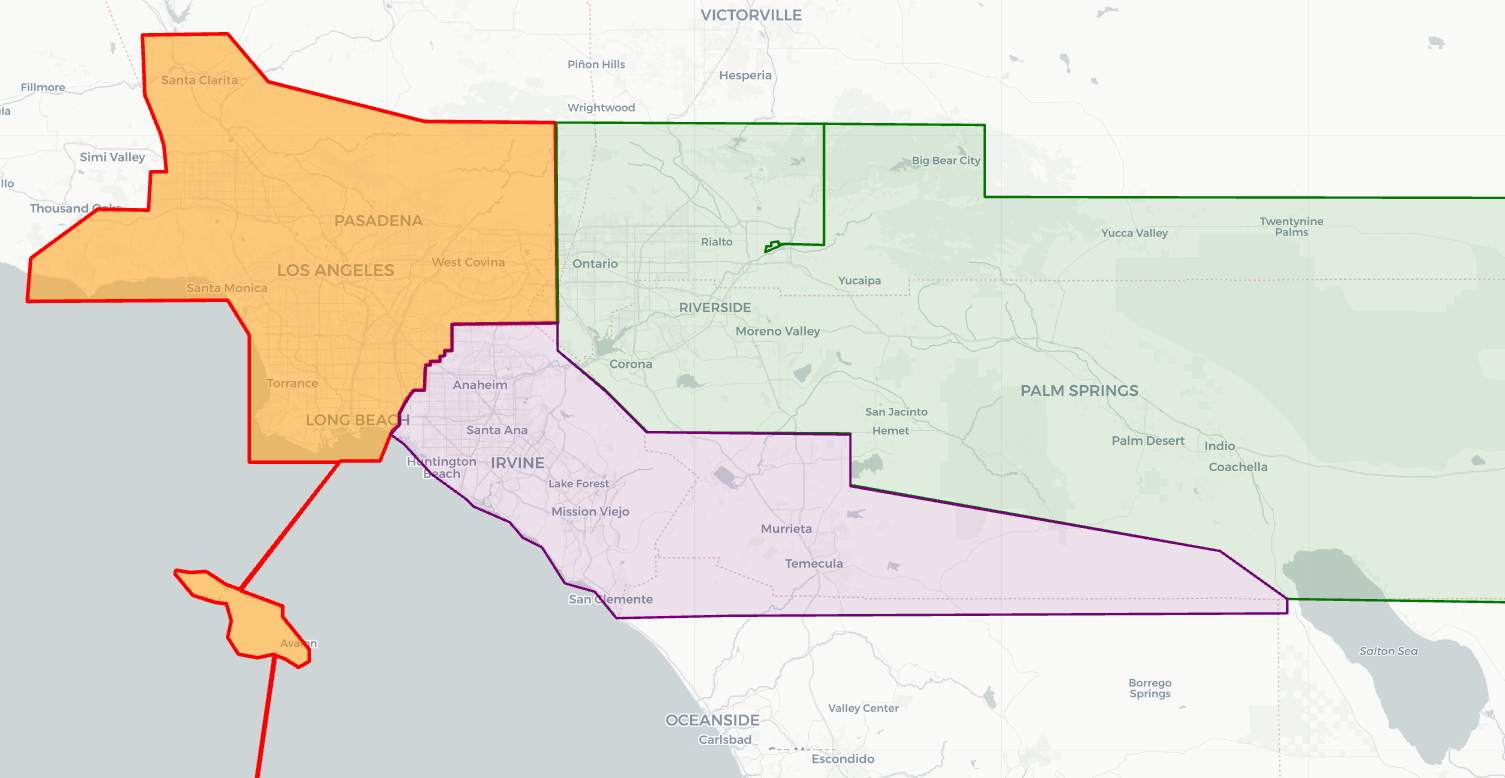}
    \caption{\centering  Illustrates the major market (orange) and the adjacent markets (green and purple).}
    \label{fig:map}
    \vspace{-1em}
\end{figure}

\subsection{Primary Outcome Variables}

We aggregate trip-level data from these three target markets into a driver-week balanced panel of weekly data, comprising 75,482 drivers observed over 25 weeks—from the week of November 7–13, 2023 to the week of April 29–May 5, 2024. (As we discuss later, we create a similar panel for the corresponding time window in 2022-23 as well.) We construct driver and trip outcome measures across four categories: driver activity, utilization, earnings, and service quality. Variables in the driver activity category include the total number of trips and sessions completed by a driver in a given week, as well as the number of hours spent online and the total miles driven. The primary supply dimension of interest is the number of hours spent online. Following \cite{Hall_Horton_Knoepfle_2023}, we define driver hours actively worked as the total time a driver was “online” on the platform, encompassing time spent on trips, time waiting for rides, and time spent en route to pick up riders \footnote{In \cite{Hall_Horton_Knoepfle_2023}, driver working time includes not only the duration spent on trips and en route to pick-ups, but also the time drivers are available to receive dispatch requests. However, due to data limitations, we do not observe the exact time when a driver becomes available or is ready to be dispatched. As a result, we define driver working time more narrowly—as the time elapsed between trip acceptance and rider drop-off, which yields a different utilization rate compared to \cite{Hall_Horton_Knoepfle_2023}'s definition.}.  Variables in the utilization category reflect how efficiently a driver converts time spent on the platform into productive activity. We measure this through the share of online hours spent actively engaged in trips (i.e., driving to or with a passenger), the number of trips per session, and trips completed per hour online. The main utilization dimension that we focus on in the analysis is the \textit{utilization rate}, which is defined as the ratio of actual driving hours with passengers and the drivers actively working.  

Variables in the earnings category capture monetary outcomes for drivers. We include average hourly earnings, earnings per trip, total tips received per week and the percentage of tips from total earnings. These measures reflect both platform-level compensation structures and rider’s tipping behavior, and are key indicators of economic incentives for drivers. We focus on the driver's hourly earning rate as the key metric of this category following the literature (\cite{Hall_Horton_Knoepfle_2023, chen2019value}). 

Finally, the service quality category of variables includes driver cancellation rate, the average rider rating received, and the average rider wait time. The primary variable in this group is \textit{wait time}, the time lapsed from requesting a ride until actually starting a trip (\cite{buchholz2022spatial}). 

The key variables and their corresponding summary statistics are presented in Table \ref{tab:summary_var} and \ref{tab:summary_stats}, respectively. The statistics are provided at driver-week level. The dependent variables we report on in the main body of the paper are in bold. We report on the other dependent variables in the Appendix. 

\begin{table}[ht]
\centering
\small 
\renewcommand{\arraystretch}{1.2} 
\begin{tabular}{p{3.5cm} p{3.6cm} p{7.9cm}}
\toprule
\textbf{Category} & \textbf{Variables} & \textbf{Definition} \\
\midrule

\multirow{4}{=}{\textit{Driver Activities} } 
& \textbf{num\_trip} & Number of trips driver $i$ has completed in week $t$ \\
& num\_session & Number of sessions driver $i$ completed in week $t$ \\
& \textbf{num\_hour} & Total driving hours online (from accepting to drop-off) for driver $i$ in week $t$ \\
& num\_mile & Number of miles driven by driver $i$ in week $t$ \\
\midrule

\multirow{4}{=}{\textit{Driver Utilization}} 
& trip\_hour & Number of hours with passengers provided by driver $i$ in week $t$ \\
& \textbf{ave\_utilization} & Fraction of hours with passengers (passenger hours / total active hours) for driver $i$ \\
& ave\_dur\_per\_session & Average active driving hours per session for driver $i$ \\
& ave\_n\_trip\_per\_hour & Average number of trips per active driving hour for driver $i$ \\
\midrule

\multirow{4}{=}{\textit{Driver Earning}} 
& \textbf{hourly\_earning} & Average earnings per active driving hour for driver $i$ in week $t$ \\
& earning\_per\_ride & Average earnings per ride for driver $i$ in week $t$ \\
& tips & Total amount of tips driver $i$ received in week $t$ \\
& perc\_tips & Average percentage of tips per rider payment for driver $i$ in week $t$ \\
\midrule

\multirow{3}{=}{\textit{Service Quality}} 
& cancellation\_rate & Rate of cancelling an accepted trip for driver $i$ in week $t$ \\
& driver\_rating & Average rating received by driver $i$, if rated \\
& rider\_wait\_time & Average rider wait time (request to pickup, in hours) for driver $i$ in week $t$ \\
\midrule

\textbf{Key Independent Variable} 
& enter\_treatment & A dummy indicating whether driver $i$ entered treatment market in week $t$ (1 if yes, 0 otherwise) \\
\bottomrule
\end{tabular}
\caption{Summary of Key Variables for Driver-week Level Analysis}
\label{tab:summary_var}
\end{table}

\begin{table}[ht]
\centering
\small 
\renewcommand{\arraystretch}{1.2} 
\begin{tabular}{p{3.5cm} p{2.7cm} p{1.5cm} p{1.5cm} p{1.5cm} p{1.5cm} p{1.5cm}}
\toprule
\textbf{Category} & \textbf{Variable} & \textbf{Obs.} & \textbf{Mean} & \textbf{SD} & \textbf{Min} & \textbf{Max} \\
\midrule
\multicolumn{7}{l}{\textit{Key Dependent Variables}} \\
\addlinespace
\textbf{Driver Quantities} & num\_trip & 1,887,050 & 12.478 & 22.618 & 0 & 562.000 \\
& num\_session & 1,887,050 & 2.519 & 3.842 & 0 & 37.000 \\
& num\_hour & 1,887,050 & 5.202 & 9.027 & 0 & 98.406 \\
& num\_mile & 1,887,050 & 90.017 & 157.556 & 0 & 1,710.242 \\

\textbf{Driver Utilization} & trip\_duration & 1,887,050 & 3.792 & 6.638 & 0 & 70.070 \\
& ave\_utilization & 1,887,050 & 0.338 & 0.367 & 0 & 0.999 \\
& ave\_dur\_per\_session & 1,887,050 & 0.946 & 1.492 & 0 & 49.217 \\
& ave\_n\_trip\_per\_hour & 1,887,050 & 1.156 & 1.387 & 0 & 57.143 \\

\textbf{Driver Earning}
& hourly\_earning & 1,887,048 & 15.399 & 20.042 & 0 & 2,153.995 \\
& earning\_per\_ride & 1,887,048 & 7.121 & 10.621 & 0 & 577.539 \\
& tips & 1,887,050 & 14.488 & 28.457 & 0 & 1,799.010 \\
& perc\_tips & 1,887,050 & 0.026 & 0.044 & 0 & 3.880 \\

\textbf{Service Quality}
& weekly\_cancel\_rate & 1,887,050 & 0.032 & 0.067 & 0 & 1.000 \\
& driver\_rating & 879,278 & 4.975 & 0.115 & 1.000 & 5.000 \\
& actual\_wait\_time & 1,887,050 & 0.072 & 0.088 & 0 & 18.403 \\

\addlinespace
\multicolumn{7}{l}{\textit{Key Independent Variable}} \\
\addlinespace
\textbf{Treatment Variable} & Enter\_treatment & 1,887,050 & 0.253 & 0.433 & 0 & 1.000 \\
\bottomrule
\end{tabular}
\caption{Summary Statistics of the 15 Outcome Measures (at driver-week level)}
\label{tab:summary_stats}
\end{table}

\subsection{Construction of the Treatment Indicator}

The intervention was launched on Tuesday, February 6, 2024; accordingly, we designate the week of February 5–11, 2024 as week(0). We define a driver as treated in week(t) if they completed at least one trip in the LAX area during that week, and had not completed a trip in the LAX area after February 5, 2024 but prior to week(t)  Aligned with the platform’s earnings payment cycle for drivers, we define each week from Monday to Sunday. The 12 weeks prior to treatment—spanning November 13–19, 2023 to January 30–February 5, 2024—are labeled week (-12) through week (-1). The 12 post-treatment weeks, from February 12–18, 2024 to April 29–May 5, 2024, are labeled week (1) through week (12). Each week beginning from week (0) thus defines a cohort of drivers who enter the treatment group in that week. By the end of week(t), the treated group includes all drivers who entered treatment in week(0) through week(t). In total, 45,205 drivers entered the treatment group during the observation period. The remaining 30,277 drivers—those who did not complete any trips in the LAX area between February 5, 2024 and May 5, 2024—are in the “never-treated” group. Figure 3 illustrates the weekly distribution of drivers across the treatment cohorts and the never-treated group over the full 25-week study period.

\begin{figure}[ht]
    \centering
    \includegraphics[width=0.8\linewidth]{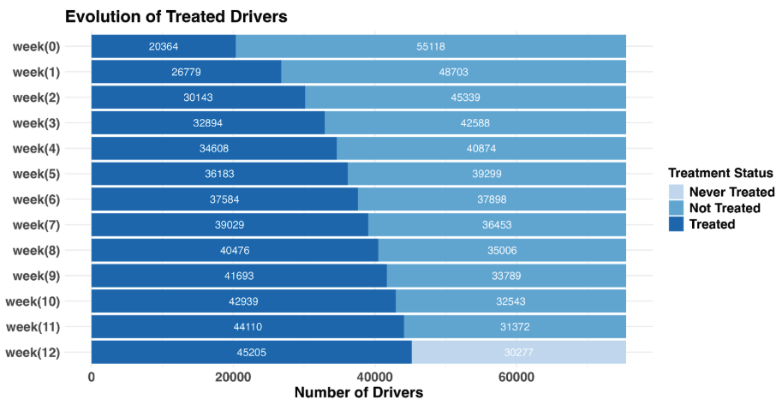}
    \caption{\centering  Summarizes the sizes of the treated set of drivers as their staggered entry into their treatment cohorts progresses over the 13 weeks of treatment.}
    \label{fig:evolution}
    \vspace{-1em}
\end{figure}

\section{Measuring Behavioral and Outcome Changes}

While our analysis is made feasible by the natural experiment described earlier, we now describe how we address the empirical and identification challenges that arise due to the nature of the policy rollout. We employ a sequence of estimation strategies, each designed to mitigate specific statistical or econometric issues. 

As described earlier, different cohorts of drivers enter the LAX market for the first time in each of 13 different weeks following the policy launch. Owing to this staggered rollout, treatment timing may not be random—drivers with certain characteristics—such as stronger engagement or greater proximity to the treatment market—may be more likely to transition into treatment sooner. For example, a driver primarily based in adjacent markets (e.g., OCX or SBD) might become treated by dropping a passenger off in LAX in a particular week, and may resemble earlier-treated drivers more than those who never enter the treated market. In such cases, differences in outcomes could reflect pre-existing characteristics rather than the effect of the policy itself. 

\begin{figure}[ht]
    \centering
    \includegraphics[width=0.8\linewidth]{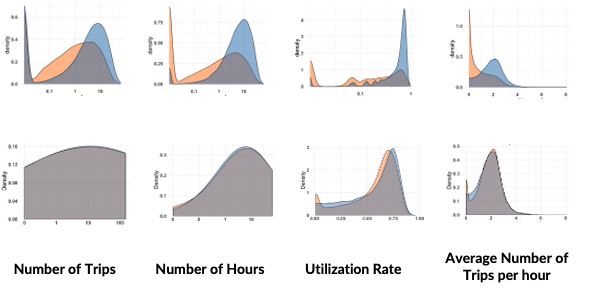}
    \caption{Illustrates the distribution of some key outcome measures for the treated (light blue) and not-yet-treated (light orange) groups for the week(0) cohort in the pre-treatment 12-week period. The grey area indicates overlap between the distributions. The top row illustrates the contrast between the groups along these measures prior to our application of inverse propensity weighting (IPW), discussed in Section 4.2, and the bottom row shows that these distributions are largely indistinguishable following the use of IPW. Table 3 in Online Appendix summarizes how, following the application of IPW, the averages of the characteristics of these groups are also comparable. 
}
    \label{fig:ipw}
    \vspace{-1em}
\end{figure}

As illustrated in Figure \ref{fig:ipw}, there is indeed an imbalance (as measured by a set of covariates corresponding to the \textit{pre-treatment values} of our primary outcome variables) between treated drivers and control drivers. The column “Treated” reports the pre-treatment period average covariate values for all drivers in the treatment cohorts. The column“Never-treated drivers” reports the corresponding pre-treatment period average value of the covariate for the outcome values, and  although not illustrated, covariate values, across the treated groups and their corresponding not-yet-treated group. While we illustrate this for one cohort, the differences persist across all 13 cohorts. 

Next, the effect of entering the treatment group is likely to be heterogeneous – it may vary across different driver groups or exhibit temporal dynamics across different time periods. It is generally accepted that standard two-way fixed effect (TWFE) models are not reliable in contexts that contain more than two groups and two time periods in part because TWFE estimators may use already-treated units as implicit controls for newly-treated units—contaminating the estimation sample and conflating temporal dynamics with causal effects. Moreover, assuming homogeneous treatment effects across cohorts can distort both the post-treatment estimates and the pre-trend diagnostics (\cite{callaway2021difference}). 

These aforementioned empirical challenges can be addressed using the estimators for staggered difference-in-differences designs proposed by \cite{callaway2021difference}, in conjunction with approaches to construct a valid estimation sample of drivers. We describe our approach in the following section.

\subsection{Estimation Framework and Identification Strategies}

We begin by estimating treatment-on-the-treated (ATT) effects using these state-of-the-art staggered difference-in-difference estimators that account for treatment effect heterogeneity. We then refine our sample along several behavioral and geographic dimensions to address endogeneity concerns accordingly. Finally, we complement these ATT analyses with an intent-to-treat (ITT) framework using a within-subject panel across two years, which leverages the fixed timing of the policy launch and holds the population constant—thus offering a clean benchmark for policy evaluation under weaker behavioral assumptions.

\subsubsection{Callaway and Sant’Anna (2021) Estimator for ATT Estimation}

Given that the objective is to estimate the treatment effect of the intervention on driver behavior, our  baseline analysis implements the estimator proposed by \cite{callaway2021difference}, which recovers dynamic cohort-specific treatment effects while addressing biases arising from comparing early-treated and already-treated units. 

Given the timing and design of the rollout, wherein drivers enter the treated market in different weeks, valid control groups for each treated cohort must be carefully chosen. In our setting, the standard unconditional parallel trends assumption is likely violated: as shown in Figure 3 and Table C2, drivers who eventually enter a treatment cohort differ from those who never do. For instance, treated drivers exhibit substantially higher levels of pretreatment activity across metrics such as total hours, miles, and number of trips. To address this, we follow the recommended approach from \cite{callaway2021difference} and \cite{santanna2020doubly} by using inverse probability weighting (IPW) to construct valid control groups composed of not-yet-treated drivers. During each period, the control group includes drivers who have not yet received treatment, regardless of whether they will be treated later. This strategy ensures that comparisons are made against units who are observationally similar and drawn from the same eventual treatment pool, rather than structurally different never-treated units. Appendix B provides formal details of the estimation framework, and Appendix C reports the covariate balance and propensity score distributions before and after weighting. This ATT estimation serves as the foundation for our subsequent analyses. 

In the next sections, we explore how the identification strategy can be further strengthened through restricted estimation samples and alternative designs that address additional sources of endogeneity.

\subsubsection{Models 1: Restricting on geographic proximity and pretreatment exposure
}

\begin{figure}[ht]
    \centering
    \includegraphics[width=0.8\linewidth]{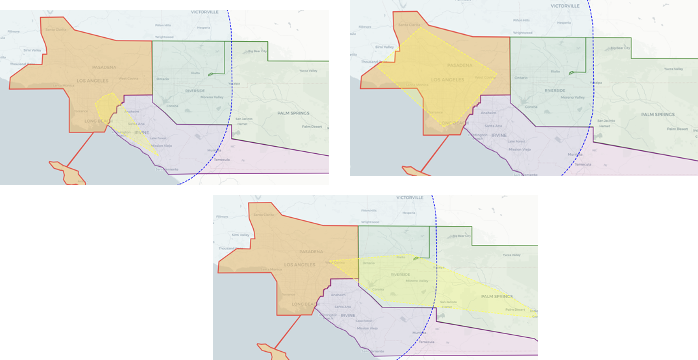}
    \caption{Illustration of driver selection criteria for the estimation samples used in Model 1. The light yellow polygon demonstrates a convex hull for a particular driver, and the blue dashed line indicates a 50km border outside the LAX market boundary. (a) Example of a driver included in the Model 1 estimation sample. (b) Example of a driver included in Model 1, whose pretreatment convex hull falls within a 50 km buffer zone around the LAX market boundary. (c) Example of an excluded driver, whose pretreatment convex hull spans an area beyond the 50 km buffer zone.}
    \label{fig:restriction}
    \vspace{-1em}
\end{figure}

Drivers whose pre-treatment region of activity is located farther from the LAX boundary may incur greater costs or frictions to access the treated market compared to drivers already operating nearby. To mitigate this form of endogenous selection, we construct an estimation sample that balances treated and untreated drivers based on their baseline geographic exposure and engagement. We summarize each driver’s historical activity space by constructing the convex hull of their pretreatment window pickup locations. The convex hull is the smallest convex polygon that contains all observed pickup points for a driver. Each polygon captures the outermost boundary of that driver’s activity space, providing a flexible but interpretable measure of geographic range that does not assume a fixed service radius. Intuitively, it represents the geographic footprint of a driver’s service area prior to the rollout of the intervention. This measure allows us to distinguish between drivers with different baseline geographic activity regions. For instance, drivers whose convex hull is tightly concentrated around the LAX boundary are likely to be more naturally exposed to the treated market, whereas drivers with a larger or more dispersed convex hull may require substantially more effort to enter the treated market. (See Figure \ref{fig:restriction}(c).)

To improve comparability, we restrict our sample to drivers whose pretreatment convex hull lies entirely within 50 kilometers of the LAX market boundary, as this distance reflects the typical trip radius between the LAX market and adjacent markets, thereby ensuring uniform treatment exposure following the policy rollout (See Figure \ref{fig:restriction}(a).) This also includes a smaple of drivers whose entire pretreatment activity is located within the LAX market, (See Figure \ref{fig:restriction}(b).). In addition, we require that drivers have completed at least one trip in LAX during the pretreatment period, further ensuring prior exposure.This yields 55219 drivers in the estimation sample used in Model 1. 

Together, these restrictions ensure that the treated and comparison drivers are comparable in terms of their baseline proximity and engagement with LAX, thereby reducing concerns about heterogeneous entry effort. 

\subsubsection{Model 2: Geographic Border Strategy}
Our first strategy restricts the estimation sample to drivers with similar baseline geographic proximity and pretreatment exposure to the LAX market—ensuring comparability in entry effort and historical engagement. However, even among drivers historically operating near LAX, unobserved behavioral or motivational differences could influence the timing of treatment adoption. To further mitigate concerns about endogenous entry timing, we estimate an additional set of models employing a \textit{geographic border strategy} that exploits quasi-experimental variation in treatment exposure among drivers based on their proximity to the LAX market boundary. 

\begin{figure}[ht]
    \centering
    \includegraphics[width=0.8\linewidth]{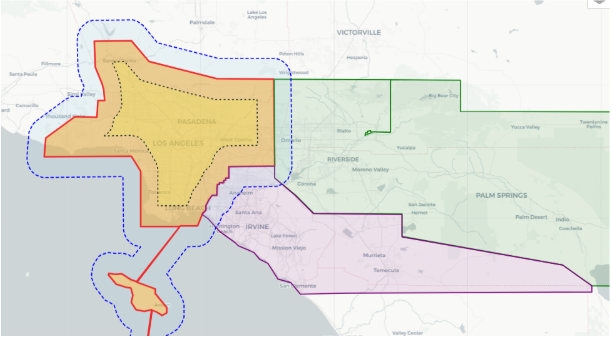}
    \caption{\centering Border strategy: the blue dotted line represents the 10km buffer zone outside of LAX, and the gray dotted line represents the 10km buffer zone inside of LAX boundary.
}
    \label{fig:border}
    \vspace{-1em}
\end{figure}

Specifically, for each post-treatment week, we compare drivers who crossed just inside the LAX boundary—thus becoming enrolled in the intervention—to those who remained just outside and were not treated. These two groups are geographically proximate, face similar market conditions, and exhibit comparable distributions of historical trip shares across the three markets (LAX, OCX, SBD) in the pre-treatment period. Crucially, the only distinction between them is whether they crossed into the treated market and completed a trip within the LAX boundary in a given week. This enables us to attribute observed outcome differences to marginal adoption of the treatment, rather than to broader differences in engagement or location. Further we restrict this sample to drivers with sufficiently similar pre-treatment driving patterns, as described in Appendix C. 

To implement this strategy, we construct a 10-kilometer buffer zone outside and around the LAX market. Treated drivers are defined as those who completed at least one trip within the inner (treated market) buffer zone in a given post-policy week. Control drivers are defined as those who, in the same week, completed at least one trip within the outer (untreated market) buffer zone but did not cross into the LAX market. (See Figure \ref{fig:border}.) This local comparison design ensures that both groups are subject to nearly identical geographic and economic conditions, mitigating concerns that treatment or its timing reflects unobserved differences in motivation, awareness, or effort that manifest as choices to drive or not drive into the LAX market. Because we construct this comparison week by week, we are able to account for dynamic entry patterns and trace out the effect of the policy on the margin of treatment adoption. 

\subsubsection{Model 3: Using Prior Year Outcomes for Treated Drivers as the Control}

Our fourth model uses a sample that further lowers any unobserved heterogeneity across the treatment and control groups by constructing a stricter balanced panel that exclusively comprises a subset of treated drivers and allows for comparison of the same individuals across years.

We restrict the sample to drivers who completed trips in both the pre- and post-treatment windows during the current year (2023–2024) and the prior year (2022–2023). This eliminates drivers who dropped out, were newly onboarded, or were only intermittently active, ensuring that our analysis compares the same drivers across years in a consistent manner. Within this balanced panel, we further restrict attention to drivers whose pre-treatment activity levels are comparable across the two years. This rules out drivers who fundamentally changed their engagement with the platform (e.g., shifting from full-time to part-time driving), thereby reducing the risk that our estimates are confounded by strong pre-existing behavioral trends. 

To formally ensure balance, we implement propensity score matching. Specifically, restricting ourselves to only treated drivers, we (i) match on driver ID, ensuring we track the same individuals across years; (ii) match on pre-treatment covariates including hours worked, miles driven, trips completed, sessions, and earnings. (The data for estimating the propensity scores thus has a dependent variable value of 1 for the 2023-24 period, and a dependent variable value of 0 for the 2022-23 period.). Covariate balance after matching is demonstrated in Appendix C. This procedure ensures that we compare drivers with stable behavioral profiles across both years, mitigating concerns that our results are driven by selection or compositional changes. This procedure yields a sample of 5,055 drivers.

\subsubsection{Model 4: One-shot Intervention }
Finally, building on the preceding analysis of the Average Treatment Effect on the Treated (ATT), which captures behavioral responses among treated drivers, we next adopt an Intent-to-Treat (ITT) framework to estimate the overall policy effect at the market level. This approach treats the earnings guarantee and transparency policy as a one-shot intervention and identifies its average impact on all potentially exposed drivers, regardless of their actual entry timing, intensity of participation, or even awareness of the policy. 

Much like Model 3, the sample is restricted to treated drivers and the comparison is across behaviors before and after the intervention and across corresponding calendar weeks in the current year (2023–2024) and prior year (2022–2023), allowing us to isolate aggregate shifts in outcomes attributable to the intervention. Starting with the Model 4 sample of 5,055 drivers, we further select drivers based on the geographic patterns of their activity in the pre-treatment window. Specifically, we restrict the sample to drivers who completed all pretreatment trips in the LAX market across both years and who were treated in week(0), yielding a sample of 1,335 drivers. These criteria ensure that our estimation sample consists of consistently active drivers with direct exposure to the treated market from the very beginning of the policy. Considering the  intervention as a one-shot treatment, we employ a standard difference-in-difference approach that uses observations in 2022-23 as a control group, and those in 2023-24 as a treated group. The equation can be expressed as: 
$$
\text {Outcome}_{\text{iyt }}=\beta_1 \text { treated }_y+\beta_2 \text { post }_t+\beta_3 \text { treated }_y * \text { post }_t+d_{i y t}+\alpha_i+\lambda_t,
$$

where $i$ indexes drivers; $t$ indexes the week of the year, and $y$ indexes the two-year period. treated ${ }_y$ indicates 1 if $y=2023 \_2024$ (the treatment year), and 0 otherwise (the control year); post $_t$ indicates 1 if the week of the year is later than February 6,2024 , corresponding to the policy implemented date, and 0 otherwise. We also control for driver fixed effects $\alpha_i$, and week-of-the-year fixed effects $\lambda_t$. Because the demand conditions differ between 2022-2023 and 2023-2024, we also present the demand distribution for both periods in Appendix G. To further isolate the treatment effect from market expansion in which underlying demand changes, we construct a driver-week-year level control for demand exposure, denoted as $d_{i y t}$, which is discussed further in Appendix G. Our models are summarized in Table \ref{tab:summary_methods}.

\begin{table}[H]
\centering
\caption{Summary of Identification Strategies and Estimation Samples}
\begin{tabular}{p{5cm} p{6cm} p{3cm}}
\toprule
\textbf{Identification Issue} & \textbf{Addressed By} & \textbf{Estimation Sample} \\
\midrule
Heterogeneous treatment effect & Adopting CSA Estimator & -- \\
Engagement level heterogeneity & IPW control for pretreatment engagement level & -- \\
Entry effort and geographic heterogeneity & Restricting driver proximity to LAX boundary & Model 1, 4 \\
Endogenous entry time & Adopting geographic border strategy & Model 2, 4 \\
Sample composition bias & Using prior year data as control group & Model 3, 4 \\
\bottomrule
\label{tab:summary_methods}
\end{tabular}
\end{table}

\section{Empirical Results}
\subsection{Direct Effects on Drivers’ Behavior}

As discussed in Section 3, we construct fifteen distinct outcome variables across four dimensions to capture changes in driver behavior and performance (see Table \ref{tab:summary_var} for variable definitions and Table \ref{tab:summary_stats} for summary statistics). We report and discuss the findings for one primary outcome from each dimension: hours online (driver activity), hourly earnings (earnings), utilization rate (utilization), and rider wait time (service quality). Results for the other outcome variables are reported in Appendix C.

\begin{table}[ht]
\centering
\begin{tabular}{lcccc}
\toprule
\textbf{Dependent Variable} & \textbf{Model 1} & \textbf{Model 2} & \textbf{Model 3} & \textbf{Model 4} \\
\midrule
\multicolumn{5}{c}{\textit{Overall Treatment Effect}} \\
\addlinespace
log(num\_hours) & 0.2874*** & 0.1775*** & 0.1964*** & 0.0823*** \\
               & (0.00057) & (0.01511) & (0.0316)  & (0.0239)  \\
log(num\_trips) & 0.379***  & 0.2298*** & 0.2885*** & 0.1223*** \\
               & (0.0076)  & (0.0220)  & (0.04477) & (0.0311)  \\
utilization rate & 0.1189*** & 0.0667*** & 0.1127*** & 0.0386*** \\
               & (0.0019)  & (0.00493) & (0.01056) & (0.0073)  \\
log(hourly\_earnings) & 0.5595*** & 0.3007*** & 0.5107*** & 0.1663*** \\
               & (0.0088)  & (0.0251)  & (0.0492)  & (0.0351)  \\
\midrule
Driver FE      & Yes       & Yes       & Yes       & Yes       \\
Week FE        & Yes       & Yes       & Yes       & Yes       \\
Drivers        & 55,219    & 9,040     & 5,055     & 1,335     \\
Observations   & 1,380,475 & 226,000   & 252,750   & 66,750    \\
\bottomrule
\end{tabular}
\small
\\ \textit{Notes:} Standard errors in parentheses. *** $p<0.01$, ** $p<0.05$, * $p<0.1$

\caption{Overall Treatment Effect: Summarizes overall treatment effect on four key outcome measures. The coefficients reported for Models 1 through 4 represent the average treatment effect across all 13 cohorts. The control group used here and in what follows (unless explicitly specified otherwise) is the not-yet-treated set of drivers with outcome measures weighted using inverse propensity score weighting. We use driver-level and week-level fixed effects.}
\label{tab:overall_treatment_effect}
\end{table}

We also provide estimates from all estimation samples we have constructed, as described in Section 4.2. As reported in Table \ref{tab:overall_treatment_effect}, there is a significant causal effect that the intervention has \textit{on all four outcome measures}. The results from each of our four models  provide clear and striking evidence that the driver response to the earnings guarantee and transparency was positive and significant.  Table 4 summarizes the average treatment effects across our four key outcomes measures. When converted into percentage effects, these results indicate, depending on the model, an increase in supply (as measured by driver hours) of between 8.6\% and 33.3\%\footnote{Some of the dependent variables in Table \ref{tab:overall_treatment_effect} have been log-transformed. Thus, converting coefficient values into percentages is according to the formula \% change = [exp(coefficient) - 1] * 100\%.  
So, for example, a coefficient value of 0 would reflect no change, and a coefficient value of 0.1 would reflect a change of exp(0.1) - 1 = [1.105 - 1] * 100\% = 10.5\% in the post-treatment period relative to the appropriate control}. The number of hours a driver chooses to bring to a platform is the most direct and immediate measure of driver reaction, and the results clearly show that the intervention was received favorably by drivers, reflecting positive sentiment towards the intervention and trust in the platform.

Furthermore, the increase in the number of trips completed increased by between 13\% and 34.8\% — note that this reflects an increase in trips completed that is higher than the corresponding increase in hours across each of the five models, which shows that both existing and additional hours were converted into trips more efficiently than in the pre-treatment window, a point reinforced by the increases in utilization rate that we also report on. We probe deeper into the increases in production efficiency in a later section. This indicates that the supply increases disproportionately “grew the pie,” and this expansion coupled with the earnings guarantee raised the hourly earnings received by treated drivers by between 18.1\% and 80.6\%. We further find that hourly revenue (platform’s take) generated from treated drivers also increased by between 10.5\% and 63.2\%. 

\begin{figure}[ht]
    \centering
    \includegraphics[width=0.8\linewidth]{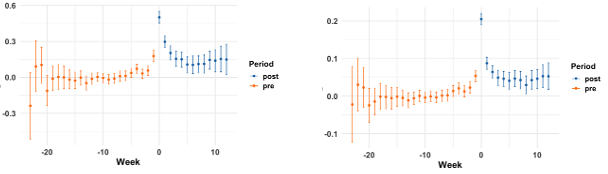}
    \caption{\centering  Illustrates the dynamic changes in the average treatment effects over time for (a) the average number of working hours a driver supplied (left) and the utilization rate (right). 
}
    \label{fig:event_study}
    \vspace{-1em}
\end{figure}

Figure \ref{fig:event_study} illustrates some dynamics of the treatment effect over time. While uniformly positive in the weeks following the study, the impact was higher initially, and while remaining positive, tapered off gradually over time. The newness of the policy (and perhaps the excitement associated with the earnings guarantee and enhanced transparency) leads to a higher treatment effect on many driver outcomes variables in the week immediately following entering the treatment group. While this initial excitement fades, the effect remains. 

We also conduct a series of placebo tests to assess the robustness of our findings by replicating our analysis for the same time window in 2022-23, which further increases our confidence that the effects of the intervention we identify in 2023-24 are in fact causal. The details are provided in Appendix D.  

As illustrated in Figure \ref{fig:event_study}, the treatment effects, while uniformly positive in the weeks following the study, are not uniform. The newness of the policy (and perhaps the excitement associated with the earnings guarantee and enhanced transparency) leads to a higher treatment effect on many driver outcomes variables in the week immediately following entering the treatment group. While this initial excitement fades, the effect remains. 

\subsection{Heterogeneity in Treatment Effects}
Next, we examine sources of heterogeneity in these treatment effects. We explore heterogeneity along a few different dimensions. We decouple the effects of the earnings guarantee and earnings transparency by segmenting drivers based on their earnings stability prior to the rollout. We use an internally defined measure of “uncertainty tolerance” to examine if the earnings guarantee might have been more favorably received by drivers with more volatile pre-treatment earnings. We also uncover other groups of drivers across which treatment effects vary by using k-means clustering.  

\textbf{Earning guarantee versus earning transparency.} The economic rationale for why an increase in a driver’s effective wage rate might lead to positive behavior changes seems straightforward. Notably, however,  the earnings guarantee was non-binding. That is, a significant fraction of drivers earned 70\% or more of the rider payments even prior to the rollout of the guarantee. It seems plausible that drivers for whom the effect of the intervention was not merely greater visibility into their earnings or a sense of security that comes from having a floor of sorts on weekly earnings but also an anticipated increase in their take rate and ensuing earnings are likely to respond more favorably to the change. 

We unpack the dichotomy of the intervention in the following way. We count the number of pretreatment weeks in which a driver nevertheless earned over 70\% of net passenger payments . We then estimate separate models on two distinct sub-samples: one comprising drivers who earned more than 70\% of their associated net passenger payments in every pretreatment week (group HH), and the drivers who had at least one week in which they earned less than 70\% of their associated net passenger payments (group LH). We adopt model 1 as the basis for all heterogeneity analysis in this section (with model 2, 3, 4 as robustness check). 

A comparison of the average treatment effects across these two groups shows that although there was an increase in the number of weekly driving hours (and other outcome measures) for drivers in both groups HH and groups LH in the post-treatment period, the effect was more pronounced for the drivers in group LH. For group HH, the weekly number of driving hours has increased by 30.7\% in the post-treatment period. For group LH, the increase was 41.7\%.  The interpretation is straightforward, supporting the idea that both the earnings guarantee and earnings transparency had independent positive effects on driver choices. One might also interpret the relatively small difference in coefficients as reflecting the earnings transparency being more important than the earnings guarantee as a determinant of behavior change. 

\begin{table}[ht]
\centering
\small
\renewcommand{\arraystretch}{1.3}
\begin{tabular}{p{4cm}p{3.5cm}p{3.5cm}}
\toprule
& \textbf{Group HH} & \textbf{Group LH} \\
\midrule
\textbf{DV} & \multicolumn{2}{c}{log(num\_hours)} \\
\midrule
Overall treatment effect & 0.2679\textsuperscript{***} & 0.3488\textsuperscript{***} \\
                         & (0.0301) & (0.0062) \\
\addlinespace
Number of drivers        & 7232 & 47987 \\
Observations             & 180800 & 1199675 \\
\bottomrule
\end{tabular}
\caption{Illustrates heterogeneity in the main treatment effect for the number of hours driven across group HH (drivers whose weekly earnings  rate was consistently above 70\% of net passenger payments  for the 13 weeks preceding the rollout), and LH (drivers who experienced at least one week during the 13 weeks preceding the rollout where their earnings rate was less than 70\% of net passenger payments).}
\label{tab:het_effect1}
\end{table}

\textbf{Variation by pretreatment earnings uncertainty.}  Next, we turn to the effects that the earning guarantee might have had on uncertainty reduction and driver risk. The floor of 70\% could be perceived by drivers as reducing the uncertainty associated with the effort they expend. While it would have been ideal to examine how driver responses varied by risk tolerance, we do not have a driver covariate that measures this individual characteristic. We can, however, using the data we have, consider a specific characterization of drivers based on the extent to which there was pretreatment \textit{uncertainty} in their earnings. Assuming that drivers are risk-averse, this reduction in uncertainty would have a positive effect. 

We compute, for each driver, the variance of hourly earnings within each of the pretreatment week. This is based on the following (loose) argument of revealed preference. A driver who values earnings certainty is more likely to work in a way that ensures that the amount they earn per hour is predictable — by targeting areas with predictable demand, accepting rides offered regularly and so on. Drivers who value this kind of predictability less are likely to make choices that lead to a more volatile series of hourly earnings. We thus assume that drivers whose pretreatment earnings variance have a lower level of \textit{uncertainty tolerance}.  Specifically, we look at two extremes — we define drivers with \textit{high uncertainty tolerance} to be those whose variance of hourly income is greater than the median variance in all pretreatment weeks, and we define drivers with \textit{low uncertainty tolerance} to be those whose variance of hourly income is lower than the median variance in all pretreatment weeks — and contrast their treatment effects with the remaining drivers. 

As shown in Table \ref{tab:uncertainty_heterogeneity}, all three groups showed a significant increase in the chosen outcome measure (in this case, average number of weekly driving hours) in the post-treatment period. However, the effect is especially high for those with low uncertainty tolerance. There is no effect we can conclusively measure for the drivers with high uncertainty tolerance. While accepting that this definition of uncertainty tolerance is a little tenuous, the difference is consistent with our hypothesis that part of the value from the earning guarantee is not just in raising earnings, but in the reassurance that a driver might get from knowing that their earnings will not vary significantly.

\begin{table}[H]
\centering
\small
\renewcommand{\arraystretch}{1.2}
\begin{tabular}{lccc}
\toprule
& \textbf{High uncertainty} & \textbf{All others} & \textbf{Low uncertainty} \\
& \textbf{tolerance} & & \textbf{tolerance} \\
\midrule
\textbf{DV} & \multicolumn{3}{c}{\textbf{log(num\_hours)}} \\
\midrule
Overall treatment effect & 0.0300  & 0.2631*** & 0.3205*** \\
                         & (0.0176)  & (0.0074)   & (0.0172) \\
\midrule
Number of drivers        & 6278    & 38380    & 10561 \\
Observations             & 156950  & 959500   & 264025 \\
\bottomrule
\end{tabular}
\caption{ Illustrates heterogeneity in the main treatment effect for the number of hours driven across drivers with high uncertainty tolerance, low uncertainty tolerance levels and all other drivers. }
\label{tab:uncertainty_heterogeneity}
\end{table}

\textbf{Other heterogeneous treatment effects:} We segment drivers into different exogenously specified categories based on their patterns of driving using k-means clustering. (See Appendix D for a list of covariates we used to cluster drivers.) We identify three main clusters of drivers, loosely corresponding to drivers who drive non-luxury vehicles for a predictable number of hours every day (“full time” or FT), those who drive non-luxury vehicles occasionally (“part-time” or PT), and those who drive a luxury vehicle (LUX).  

We find that the treatment effects remain significantly positive across all outcome measures and all three driver clusters, but the magnitude of effects varies substantially by group. For instance, in the case of weekly working hours, the effect is most pronounced among full-time drivers, who exhibit an average increase approximately 26\% larger than that observed in the part-time group. The PT drivers also experience a statistically significant gain in hours worked, albeit smaller in magnitude, while LUX drivers show the smallest effect among the three groups. 

\begin{table}[H]
\centering
\small
\renewcommand{\arraystretch}{1.2}
\begin{tabular}{lccc}
\toprule
& \textbf{Luxury Driver} & \textbf{Full-time Driver} & \textbf{Part-time Driver} \\
\midrule
\multicolumn{4}{c}{\textit{log(num\_hours)}} \\
\textbf{Overall treatment effect} & 0.1288*** & 0.3779*** & 0.3099*** \\
& (0.0137) & (0.0119) & (0.008) \\
\midrule
\textbf{Number of drivers} & 6796 & 32026 & 26711 \\
\textbf{Observations} & 169900 & 800650 & 667775 \\
\bottomrule
\end{tabular}
\caption{Illustrates heterogeneity in the main treatment effect for the number of hours driven across drivers in the three clusters revealed by k-means clustering}
\label{tab:driver_type_heterogeneity}
\end{table}

\subsection{A Sample of Strategic Responses}

While the changes in choices made by drivers and riders were largely aligned with platform incentives and speak to the intervention being a resounding success for Lyft, we also see some evidence of strategic responses among treated drivers to greater earnings transparency that may have led to unintended consequences. First, we compute the per-trip wait time, defined as the time difference between the request time and the pickup time. This measure increased by approximately 2\% for treated drivers, suggesting treated drivers may become more selective after receiving greater transparency about their earnings. Knowing their guaranteed or expected earnings in advance, they can afford to wait longer to evaluate whether a trip is worthwhile (for example, based on direction, distance, or location). Second, we calculate the cancellation rate of accepted trips, which rose by 0.6\% post-treatment. This is a small change, but might be reflective of the earning commitment policy reducing uncertainty and empowering drivers to make better-informed trade-offs. If they see that an accepted trip is suboptimal — because of a lower-than-expected take rate— they may cancel more readily, especially if the penalty is offset by the expectation of other high-margin trips.

The more interesting strategic responses had to do with clear shifts in the driver’s choices of activity locations geographically. We label pickup locations with above-median weekly demand intent as high-demand areas. For each driver-week, we then compute two outcome measures: the fraction of trips originating from high-demand locations and the fraction of trips with pickup times during peak hours (6–9 AM and 4–7 PM). As reported in Table \ref{tab:strategic}, treated drivers increased their share of trips during peak hours by 7\% and from high-demand regions by 17\%, relative to the pre-treatment period. This finding highlights a new economic tradeoff induced by greater transparency. Prior to the intervention, drivers had to make acceptance or positioning decisions under imperfect or asymmetric information — often unsure of their net margins. After transparency, drivers are armed with complete information of expected earnings, commissions, or share of fare. This richer information leads to more calculated, effort-optimizing strategies, such as canceling suboptimal trips if better ones are likely and concentrating supply during peak windows or in high-demand zones.

\begin{table}[H]
\centering
\renewcommand{\arraystretch}{1.3}
\small
\begin{tabular}{p{2.5cm}p{2.4cm}p{3.8cm}p{3.6cm}p{3.3cm}}
\toprule
\textbf{Dependent Variable} & \textbf{log(wait\_time)} & \textbf{log(cancellation\_rate)} & \textbf{log(frac\_Hdemand)} & \textbf{log(frac\_Phour)} \\
\midrule
Overall treatment effect & 0.0221*** & 0.0061*** & 0.1573*** & 0.0704*** \\
                         & (0.0003)  & (0.0004)  & (0.0475)   & (0.0127)   \\
\midrule
Driver FE                & Yes       & Yes       & Yes        & Yes        \\
Week FE                  & Yes       & Yes       & Yes        & Yes        \\
\midrule
Drivers                  & 55219 & 55219    & 55219    & 55219     \\
Observations             & 1380475 & 1380475 & 1380475  & 1380475  \\
\bottomrule
\end{tabular}
\caption{Illustrates the overall treatment effect in four outcome measures that captures drivers’ strategic responses.}
\label{tab:strategic}
\end{table}

\subsection{The Spillover Effects of Supply Into Demand}

Next, we explain how the driver behavior changes associated with the intervention led to demand spillovers. Our data includes three different measures of demand—rider intents, rider requests, and rider completes (explained shortly)—provided by zip code for each hour of the six-month window of our study. Note that a portion of this analysis, aggregation is by geography rather than by driver, since demand is clearly location- rather than driver-specific. 

When a platform user opens the app (and is thus considering requesting a ride), this increases the count of \textit{rider intents} by one. When a platform user makes an actual ride request after opening the app, this action increases the count of \textit{rider requests} by one. This can be thought of as a measure of the actualization of the rider intent. When a platform user takes and completes a ride, the count of \textit{rider completes} (or trips) increases by one. We also transform these metrics to analyze changes in demand based on conversion rates rather than absolute levels. 

We focus on four measures of demand—an absolute measure of rider intents, an absolute measure of rider requests, the ratio of rider completes to rider intents, and the ratio of rider completes to rider requests. We aggregate these measures into weekly zipcode-level demand measures, and compare demand outcomes for the zip codes from the major market (which we term the treated zip codes) with those from the adjacent markets (the control zip codes) before and after the intervention. To account for additional factors that could shift demand of a given zip code in a given week, we include a continuous price indicator variable $PriceIndicator_{it}$ which represents an aggregate factor reflecting levels of price in a specific area for a specific week are relative to historical averages. This helps control for factors such as local events and unusual surge pricing changes that shift demand. Our model is specified below: 
\begin{equation}
\text { demand }_{i t}=\alpha+\beta \text { TreatMarket }_i * \text { After }_t+\text { PriceIndicator }_{i t}+\eta_i+\eta_t+\varepsilon_{i t}
\end{equation}

The central finding is that there is a significant and positive increase in basic demand intent, as measured by the count of rider intents. Treated markets have seen an average increase of about 3\% in rider intents in the posttreatment period relative to the neighboring markets. This is clear evidence of a positive demand spillover caused by the intervention program. Put differently, the positive direct effects on supply spilled over into a secondary effect on demand — there is a causal effect that these increases in driver activity have on increasing the rate at which riders open their app and consider booking a ride, and the rate at which riders convert requested rides into completed rides. (For reasons of confidentiality related to the  platform’s aggregate demand levels, we cannot  provide coefficient values or other summary statistics for these results.) 

To better identify the economic mechanism by which the supply-side changes are being converted into demand-side effects, we conduct additional analyses that exploit spatial heterogeneity in the supply and demand patterns across the treated and control zip codes. We  hypothesize that the treatment effect led to a greater supply-side response in high-demand areas—marked by increased driver availability and reduced wait times—which, in turn, generated a larger increase in rider demand in these areas relative to low-demand areas. 

Specifically, we examined the data at both (1) the zip code level and (2) the H3 hexagon level to capture geographic heterogeneity at varying spatial granularities \footnote{  Zip codes provide a familiar administrative unit for policy relevance, while the hexagon-based grid allows for more consistent geographic resolution, mitigating the variance in zip code sizes. We have validated the robustness of our results across other geographic binning strategies, all of which consistently identified similar high-demand clusters.}. We aggregate data across the 489 zip codes in the three focal markets. These zip codes were then grouped into deciles based on their average weekly demand in the pretreatment period, with the top decile representing high-demand areas. Out of 48 high-demand zip codes, 42 are from the major market, and 6 are from the adjacent markets. As shown in Appendix Figure G1, the demand distribution is highly skewed. 

We then examined variation in the demand spillover across these demand strata. As reported in Table \ref{tab:heterogeneous_demand}, all demand-side measures increased significantly more in high-demand areas relative to low-demand areas. Specifically, basic demand intent rose by approximately 3.9\% more in high-demand treated areas, and the number of rider requests increased significantly by 5.9\% in high-demand treated areas compared to the rest of the area. 

\begin{table}[htbp]
\centering
\begin{tabular}{lccc}
\toprule
\textbf{Dependent Variable} & \textbf{log(intents+1)} & \textbf{log(num\_requests+1)} & \textbf{log(complete\_trips+1)} \\
\midrule
\begin{tabular}[c]{@{}c@{}}Treat\_market $\times$\\ high\_demand $\times$ after\end{tabular} 
& 0.0385* & 0.0567** & 0.0706*** \\
 & (0.0156) & (0.0175) & (0.0172) \\
\midrule
Zip Code Fixed Effects & Yes & Yes & Yes \\
Week Fixed Effects & Yes & Yes & Yes \\
S.E.: Clustered by & zip\_code & zip\_code & zip\_code \\
Observations & 12,225 & 12,225 & 12,225 \\
\bottomrule
\end{tabular}
\caption{Average Treatment Effect across Regions of Heterogeneous Demand Levels}
\label{tab:heterogeneous_demand}
\end{table}

Moreover, for higher demand areas in the treated market, the conversion rate from intent to completed trip increased by approximately 0.02 percentage points. While this may seem like a small increase, it is more significant than it might seem when one considers that there was an increase in rider intents (which makes the denominator of the conversion rate higher, and thus, all else equal, would lead to a decrease in the conversion rate, again, we suppress exact information about conversion rates for competitive confidentiality reasons.)

Additionally, we examine geographic variation in supply-side measures. Zip code areas vary widely in size—while some urban zip codes span less than a square mile, the average U.S. zip code covers 50-100 square miles. To complement our zip code-level analysis and address this heterogeneity, we adopt the H3 geospatial grid system to create uniform geographic units. The H3 system divides space into hexagonal cells at multiple resolutions (0 to 10) with a lower resolution corresponding to a coarser partition and larger area per hexagon; we use level 7 (Castillo, 2024), which approximates a one-mile radius around each pick-up location. This yields 5,915 hexagons across our markets, with each hexagon cell covering an average of 5.16 km². This finer spatial resolution enables more consistent comparisons and better identification of high- and low-demand areas.

We aggregate trip-level data to the hexagon-week level and group hexagons by their average number of completed trips per week in the pre-treatment period. We define the top ten hexagons as high-demand areas, and. Table \ref{tab:waittime_heterogeneity} presents the heterogeneous treatment effects across different demand levels. In high-demand areas, the average wait times decreased significantly by 3\%. Notably, this region-level wait time differs from the per-ride wait time reported in Section 5.3, and suggests that the effects of greater driver availability were disproportionately concentrated in high-demand areas, lowering wait times and leading to more efficient matching and higher converted demand. 

\renewcommand{\arraystretch}{1.3} 

\begin{table}[htbp]\centering
\begin{tabular}{lccc}
\hline
\textbf{Dependent Variable} & \multicolumn{3}{c}{\textbf{log(wait\_time)}} \\
\hline
(treated market) $\times$ & \textbf{-0.0338***} &  & \\
(after) \vspace{3pt}      & (0.0060)            &  & \\
(high demand) $\times$    &                     & \textbf{-0.0315***} & \\
(after) \vspace{3pt}      &                     & (0.0035)            & \\
(treated market) $\times$ &                     &                     & \textbf{-0.0284***} \\
(high demand) $\times$ (after) \vspace{3pt} &    &                     & (0.0031) \\
\hline
Hexagon Fixed Effects: & Yes & Yes & Yes \\
Week Fixed Effects:    & Yes & Yes & Yes \\
S.E. Clustered by:     & hexagon & hexagon & hexagon \\
Observations:          & 147,875 & 147,875 & 147,875 \\
\hline
\end{tabular}
\caption{Average Treatment Effect on Per-ride Wait Time across Regions of Heterogeneous Demand Levels}
\label{tab:waittime_heterogeneity}
\end{table}

\section{Ongoing Work: Could The Supply Shock Have Led To Even More Rides?}
A natural question that stems from the final findings reported in Section 5.4 is whether the intervention caused more nuanced system-wide changes in the distribution of supply across the region. We examine more granular supply distribution changes by measuring, at a hexagon level, the average weekly supply across each of the 12 post-treatment weeks. We then construct a Lorenz curve for each post-treatment week, compute from it the associated weekly supply Gini coefficient \footnote{The Gini coefficient is a widely used measure of inequality. See, for example, https://ourworldindata.org/what-is-the-gini-coefficient }, and compare these post-treatment Gini coefficients to the corresponding pre-treatment supply Gini coefficient computed based on average hexagon-level supply in the 13 pre-treatment weeks. 

\begin{figure}[ht]
    \centering
    \begin{subfigure}[b]{0.45\linewidth}
        \centering
        \includegraphics[width=\linewidth]{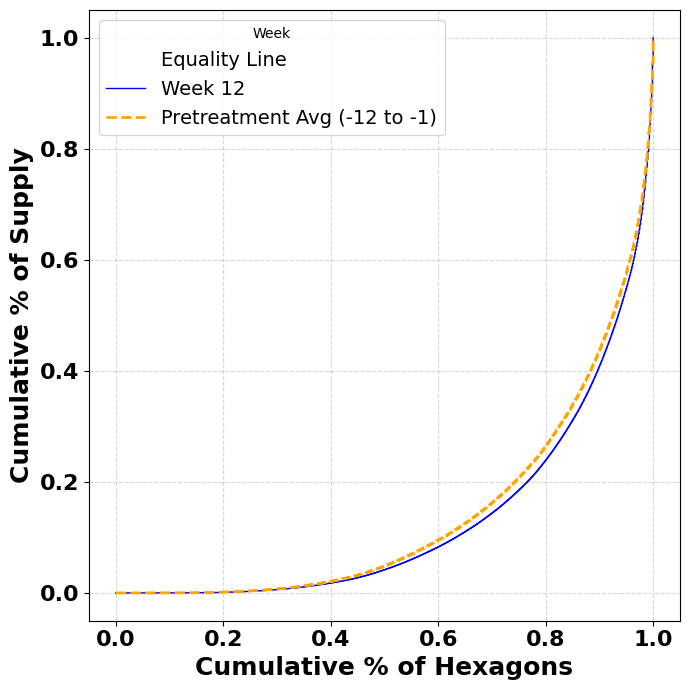}
        \caption{Lorenz Curve}
        \label{fig:lorenz-left}
    \end{subfigure}
    \hfill
    \begin{subfigure}[b]{0.5\linewidth}
        \centering
        \includegraphics[width=\linewidth]{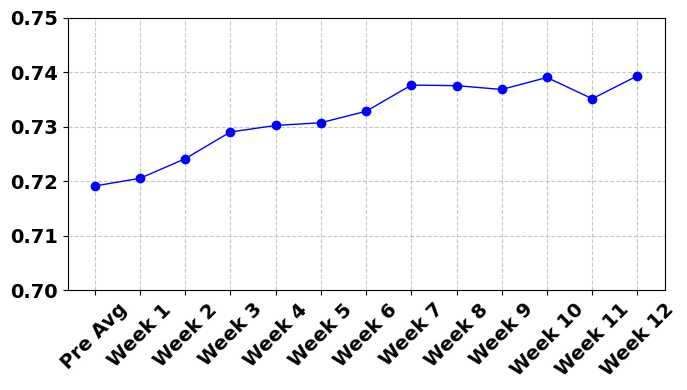}
        \caption{Gini Coefficient Over Time}
        \label{fig:lorenz-right}
    \end{subfigure}
    \caption{Illustrates the changes to the Lorenz curve (left) and the Gini coefficient of supply (right), measured based on weekly average hexagon-level supply.}
    \label{fig:lorenz}
    \vspace{-1em}
\end{figure}

Our preliminary results are illustrated in Figure \ref{fig:lorenz}. We document a gradual (albeit slight) increase in the supply Gini coefficient over time. While this analysis remains in progress, it underscores the complex interplay between different equity objectives when reshaping platform policy. While it seems natural to interpret the earnings guarantee coupled with greater transparency as a step towards greater platform-driver equity (much like one might interpret any corresponding “minimum wage” government intervention), this preliminary finding suggests that such interventions may induce unplanned changes in other metrics of inequality.

We conclude by describing ongoing work that dives deeper into the efficiency with which the supply increase induced by the intervention was “converted” into completed rides. One might think of the production system of a ridehailing platform as having two key inputs — the supply or availability of driver hours, and the demand or expressed intent of riders to take rides — that are converted by the ridehailing “production” system into completed rides \footnote{ The mapping from driver supply and rider demand into rides could be thought of as either a production function or a matching function. We choose to use the term “production function.”}. Indeed, this approach to ride production has been taken by prior work, including \cite{castillo2023matching}. Granted, we have already established that the supply increases induced by the intervention led to systemwide benefits. However, since driver supply and the rider demand must be collocated in space and in time for the system to be able to convert them into completed rides, a natural question that arises is whether the positive supply shock induced by the intervention was efficiently distributed, or whether the same intervention could have led to an even greater increase in ride production. 

A core challenge in modeling counterfactual ride production lies in the fact that driver hours are not physical quantities that can be arbitrarily reallocated across time and space. While drivers choose where to start sessions, the ensuing supply is governed largely by where the riders they pick up want to go, and new supply is generated at each drop-off location. Drivers who drop off passengers at the  LAX airport, for example, cannot be magically and instantly reassigned to serve demand in a distant region like Irvine or Santa Monica. 

Thus, rather than assuming supply is fully fungible, we adopt a more pragmatic approach that we believe constructs feasible counterfactual reallocations by varying the session \textit{starting locations} and \textit{times} of drivers. We model driver supply as a distribution over space and time that is then generated by a driver’s choice of starting locations and time slots. These distributions, which are derived from the patterns of ensuing supply that we observe in our data, capture the supply infusion induced by a driver starting at a particular location on a particular day and time. We then formulate an optimization problem that reassigns driver session starts across time-space markets to optimize total ride production, subject to constraints on total supplied sessions and hours. This behaviorally-constrained reallocation provides a flexible yet managerially relevant and realistic basis for counterfactual analysis.

\subsection{Estimating a ride production (matching) function}
We use the Cobb-Douglas production function which takes supply and demand as inputs. We estimate this production function for different spatiotemporal markets. Following prior literature (\cite{castillo2024surge}), we discretize geographic space into equal-sized hexagonal cells, each covering an area of about 36 square kilometers on average. This results in a total of 220 unique hexagons across the focal region. We construct 15 geographic areas by aggregating adjacent hexagons based on metropolitan boundaries—examples include DowntownLA and CulverCity, LAX Airport Area, and Irvine and NewportBeach region. Each weekday is divided into five time slots: morning peak (7–9 AM), midday (9 AM–5 PM), evening peak (5–8 PM), off-peak (8–9 PM), and late night (9 PM–12 AM). Each of these time slots is further intersected with the seven days of the week, from Monday through Sunday. The resulting market definition spans 15 geographic areas × 5 time slots × 7 days, yielding a total of 525 distinct markets, with the implicit assumption that the parameters of the production function are the same for the hours of day and hexagons within each “market” (\cite{castillo2023matching}). Now, within each market, we use supply and demand data at the hourly-hexagon level.  

At any point in time, an active driver is in one of three mutually exclusive operational states: (1) idle (waiting for the next match),  (2) en route (driving to a rider’s pickup location), or (3) transporting (completing the ride with a passenger on board)\footnote{ Specifically, drivers are considered to be in the idle state during the intervals between the drop-off of one trip and the acceptance of the next. We impute idle durations based on these sequential trip gaps, both within and across sessions. When a driver completes a ride and initiates a subsequent trip within the same session, we assume that they remain physically located in the drop-off hexagon during the idle period. If the trip is the final one in a session, we assume a conservative two-hour threshold for the idle period, after which the driver is considered offline.}. For each hexagon-hour pair, we aggregate the fraction of time during that hour each driver spent in that hexagon in any of the three active states (Idle, En Route, Transporting). Demand is simply the number of rider intents, which, as you may recall, captures the number of users opening the app in a given hexagon-hour.

We estimate a Cobb-Douglas production function for each market $i$, modeling ride production as a function of supply and demand inputs: 

\begin{equation}
    y_{i t}=A_{i t} D_{i t}{ }^{\alpha_i} S_{i t}{ }^{\beta_i},
\end{equation}
where \begin{itemize}
    \item $y_{i t}$ denotes the number of completed rides completed in market i , hour t
    \item $D_{i t}$ is the total demand intents in market i , hour t
    \item $S_{i t}$ is the total driver availability (supply) in market i , hour t
    \item $\alpha_i$ : demand elasticity
    \item $\beta_i$ : supply elasticity
    \item $A_{i t}$ captures market-specific residual productivity or baseline efficiency
\end{itemize}

We use data about all 52 weeks in our data set. Figure H summarizes the distribution of estimates of supply and demand elasticities across the 525 markets. The sum of the supply and demand elasticities measures the returns to scale of the ride production process—a value greater than one indicates increasing returns to scale, implying that proportionally increasing both driver supply and rider demand leads to more-than-proportional growth in the number of matched trips. On average, the markets we examine display increasing returns to scale. Also, consistent with prior work, demand elasticity is significantly higher than supply elasticity. 

\subsection{Counterfactual Analysis}
\begin{figure}[htbp]
    \centering
    \begin{minipage}{0.95\linewidth}
        \centering
        \textbf{Panel A:} Distribution of Individual Driver Sessions for 9 Drivers \\
        \vspace{0.8em}

        \includegraphics[width=0.9\linewidth]{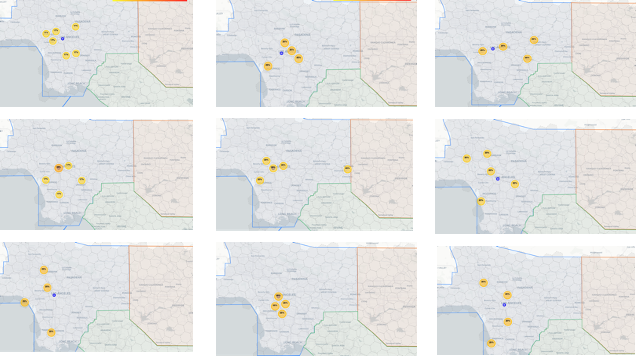}
    \end{minipage}
    
    \vspace{2em}
    
    \begin{minipage}{0.95\linewidth}
        \centering
        \textbf{Panel B:} Distribution of All Driver Sessions \\
        \vspace{0.8em}
        \includegraphics[width=0.9\linewidth]{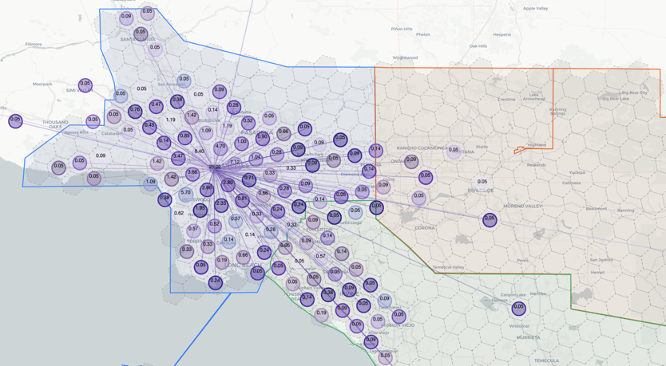}
    \end{minipage}
    
    \caption{
        \textbf{Visualizing the spatial structure of driver session trajectories.} 
        Each panel fixes the starting time to Midday on Wednesday. In both panels, the dark blue circle indicates the starting hexagon for a given driver, and each orange circle represents the ending hexagon of sessions starting from that location. The annotated percentage next to each orange circle denotes the share of sessions ending in that specific hexagon.
    }
    \label{fig:counterfactual}
    \vspace{1em}
\end{figure}

We use these estimated production functions as the basis for our  counterfactual analysis. As alluded to earlier briefly, driver hours cannot be treated like physical commodities that can be freely reallocated across space and time. Instead, we leverage the behavioral flexibility drivers exhibit in choosing their starting locations and time slots, which are more amenable to intervention and policy nudges. Figure \ref{fig:counterfactual} summarizes our approach to modeling supply changes. By examining each driver session in our data, we can calculate the fraction of ensuing supply that was received by each of the hexagons from that specific session starting location (this is illustrated for nine sample driver sessions in the first panel of the figure). By aggregating all sessions that started in a specific hexagon, we can compute, for each starting hexagon, a distribution of ensuing supply (second panel of the figure). We can now compute the distribution of overall supply that a counterfactual reallocation of driver starting locations and times will yield, which in turn is translated into a distribution over counterfactual ride production. 

To evaluate whether alternative spatial and temporal allocations of driver availability can enhance ride production, we are developing a set of heuristics that perturb each driver session’s starting location and starting time. Thus far, we have experimented with five heuristics. The first four hold the driver’s start time constant and vary the starting location in one of four different ways (1) a one‑hop strategy, which reassigns each session’s starting hexagon to one of its immediate neighbors; (2) a two‑hop strategy, which reassigns starting locations to a hexagon located within two adjacency steps; (3) a demand‑weighted two‑hop strategy, in which each session’s two‑hop neighborhood is weighted by contemporaneous demand intensity and the hexagon with the highest demand is selected; and (4) a greedy strategy, which evaluates each of the driver sessions sequentially, first finding the driver session that produces the highest marginal impact on ride production from its reallocation, fixing that session, then repeating the process. The fifth heuristic preserves the greedy strategy but instead shifts driver starting hours forward or back by an hour. 

Our preliminary results suggest that spatial shifts in driver availability yield greater increases in predicted ride production than temporal shifts. This work is ongoing, and we expect to present more details in a forthcoming paper.

\section{Concluding Remarks and Future Work}

Our results establish that Lyft’s earnings guarantee and transparency had widespread positive effects on driver engagement, with separate effects from the earnings guarantee and the enhanced transparency. Drivers increased their working hours and utilization, leading to more completed trips and higher per-hour and per-trip earnings. These effects were strongest for drivers with specific patterns of pre-treatment take rates and lower income uncertainty tolerance. Increased supply also had a positive spillover on demand, boosting rider app engagement and booking conversions in treated markets. Our results validate the platform’s bold and risky move, which made a bet that, when coupled with an earnings guarantee, greater transparency would be favorably received by drivers and lead to systemwide gains.

Our results have important policy implications, suggesting that platform-led earning floors may serve as a viable substitute for externally imposed minimum wage policies. While ride-hailing platforms have  still continued to choose pricing levels, city-mandated regulatory minimums significantly constrain the extent to which the market mechanism shapes pricing and allocation. Put differently, in such regulated markets, while market forces dictate the prices that consumers were willing to pay for ride-hailing in what remains a highly competitive market, driver earnings are constrained in parallel by regulatory minimums. Our results point to a natural alternative to minimum pay laws — let market forces dictate both consumer pricing and the division of revenues. Indeed, there are a wide variety of other platform-chosen design choices — including how consumers search, how providers and passengers are matched, and what determines provider entry and exit — which promote the efficiency and smooth functioning of this new economic form. Rather than relying on complex and often opaque formulas—such as per-minute or per-mile rates that drivers may not fully understand—an earnings guarantee simplifies earning structures while preserving platform flexibility. We hope our results inform future debates on gig worker protections by offering a hybrid solution that balances regulatory objectives with operational practicality.

As summarized in Section 6, one central focus of our current work is aimed at better understanding how future interventions that lead to positive responses might lead to greater system-wide gains. We continue to refine our production function-based counterfactual strategy, and expect to report on these findings in a follow-up paper that also yields new insights into the economics of platform-based ride production.  

We are also deepening our investigation of how platform transparency changes the strategic behavior of participants. Our finding that treated drivers exhibit a greater tendency to accept high-margin trips, particularly those concentrated in high-demand zones and peak hours,  suggests strategic optimization in response to newly revealed information, an empirical finding we wish to generalize by grounding it in a more robust theoretical framework. Traditionally, many components of ridehailing fare structure—such as dynamic pricing multipliers, surcharges, service fees, and taxes—are only partially observable to drivers. Transparency shifts this information structure by enlarging their awareness, potentially increasing trust but also reducing the platform’s informational advantage in incentive design. For example, one might think of the platform as acting as the principal in a principal-agent setting, designing algorithms and compensation structures to shape the behavior of the drivers (agents) who choose effort (hours, spatial availability) and selective behaviors (acceptance, cancellation, routing). We have not yet formalized the modeling of this behavioral change, but one structural approach suggests promising directions: embedding belief updating and selective effort into a micro-founded principal-agent model of the platform-driver interaction. This will help us evaluate whether the observed post-treatment patterns reflect rational updates to the driver’s internal payoff function, or behavioral shifts arising from newly salient trip-level attributes.

We are also working on unpacking the nature of the supply increases, and in particular, understanding what fraction of the higher hours came from drivers simply working more, and what fraction stemmed from competitive substitution. Detecting such substitution is empirically challenging, as there are no widely accepted methods to identify driver multihoming purely from observational activity data in a single platform. Nevertheless, we have developed a preliminary approach that leverages trajectory-level embeddings and contrastive representation learning to uncover latent patterns suggestive of multihoming behavior. When complete, this approach will provide a scalable, behavioral lens into the structure of driver effort and its possible reallocation across competing platforms, and opens the door to counterfactual simulations of supply-side substitution under varying platform designs.

Our research and findings offer several important managerial implications. They validate a bold platform decision: the introduction of earnings guarantees and greater transparency was a risky move that required trust from drivers. The observed behavioral changes—more hours worked, higher earnings, and shifts in strategic engagement—indicate that drivers responded positively, and provide new strategic insight to other platforms considering the relationship between building or maintaining participant trust and  their design choices. What is especially striking is that these benefits did not come at the cost of lower platform earnings or inferior rider experiences. Rather, the intervention grew the pie in multiple ways and led to a win-win-win outcome across platform stakeholders. Our ongoing work also provides a new approach to predicting platform multihoming. Our counterfactual experiments will provide actionable managerial insights by identifying when and where additional supply is most productive, guiding platforms towards fine-tuning their incentive schemes to nudge drivers toward underserved time slots and locations, and allowing the platform to allocate supply more efficiently, improving rider conversion without necessarily increasing total driver hours.

\clearpage

\bibliographystyle{informs2014} 
\bibliography{ref} 

\clearpage

\section{Appendix}
\subsection{Appendix A: Justification of Selected Markets }

We focus our analysis on the Los Angeles (LA) metropolitan area for several reasons. First, LA offers a diverse and spatially expansive ride-hailing ecosystem, with a rich mix of urban cores (e.g., Downtown LA), dense commercial corridors (e.g., Hollywood, Santa Monica), suburban communities (e.g., Orange County, Pasadena), and key transit hubs (e.g., LAX Airport). This geographic heterogeneity provides meaningful cross-market variation in supply, demand, and operational frictions, making it a robust setting for measuring elasticity and policy responsiveness across market types. Second, compared to other major cities like San Francisco or New York, LA has weaker public transit infrastructure and greater car dependency, making ride-hailing a more essential service rather than a marginal or convenience-driven mode. This allows us to more clearly isolate the causal effects of platform design changes on ride-hailing behavior, unconfounded by rapid substitution toward or away from subways or commuter rail.
Third, The San Francisco Bay Area presents a complex and fragmented airport topology, which poses challenges for isolating policy impacts on platform-mediated ride-hailing. The region is served by three major commercial airports within a relatively compact geographic radius. In contrast, Los Angeles is dominated by a single, centralized international hub—LAX, which accounts for a substantial share of airport-related ride-hailing demand in the region. While there are secondary airports (e.g., Burbank, Long Beach, Ontario), LAX remains the clear anchor for long-haul air travel and thus concentrates both the rider pool and driver strategies. Finally, New York City poses unique challenges due to its distinct regulatory regime, including medallion systems, ride quotas, and mandated earnings floors—all of which introduce strong institutional confounds. The regulatory structure in NYC differs markedly from the rest of the U.S., limiting external validity. In contrast, LA operates under more typical platform governance and labor flexibility, making it a better representation of platform labor dynamics in most U.S. markets.

\subsection{Appendix B: The Callaway and Sant’Anna (2021) Framework
}
We adapt the staggered difference-in-differences framework proposed by \cite{callaway2021difference}, hereafter referred to as “CSA,” to estimate average treatment effects of the earning guarantee and earning transparency policy on treated drivers.  The CSA approach first estimates cohort- and time-specific average treatment effects on the treated observations, using two-period/two-group difference-in-difference estimators and then aggregates them, weighting estimates by the size of each treatment cohort, to produce summary treatment effect estimates. 
The use of this estimator and associated strategy for constructing control groups addresses the econometric issues we highlighted earlier that are associated with a simpler difference-in-differences approach in the following ways. First, it accounts for the temporal variations in treatment timing across drivers. Second, it addresses the heterogeneity of drivers when constructing a comparable control group using untreated and not-yet-treated drivers and hence is not biased by time-varying treatment effects. Moreover, it allows us to estimate not only the overall treatment effect but also the dynamic treatment effect based on the length of the exposure to the treatment, doing so by aggregating drivers that entered into treatment at different weeks into different groups.

The CSA estimates an average treatment effect on the treated for each cohort. Denote $G_{i g}$ as a binary indicator that equals to one if a given driver $i$ was first treated at time $g \in\{0, \ldots, 12\}$ (thus, if driver $i$ belongs to cohort $g$ ). Denote $Y_{i t}$ as the potential outcome measure at time $t$ for a given driver $i$. Specifically, let $Y_{i t}(0)$ denote the potential outcome for driver $i$ at time $t$ were this driver untreated. Let $Y_{i t}(g)$ denote the potential outcome for driver $i$ at time $t$ if she entered the treatment group at time $g$. The group- time- specific average treatment effect at week $t(t \geq g)$ is then defined as follows:

\begin{equation}
\operatorname{ATT}(g, t)=E\left(Y_{i t}(g)-Y_{i t}(0) \mid G_{i g}=1\right)
\end{equation}

Following CSA, there are two identifying assumptions for $\operatorname{ATT}(g, t)$ :

Assumption 1: No anticipation of treatment.
\begin{equation}
    Y_{i t}(g)=Y_{i t}(0) \text { for all } t<g
\end{equation}

Assumption 2: Conditional Parallel Trends
\begin{equation}
    E\left[Y_{i t}(0)-Y_{i t-1}(0) \mid G_{i g}=1\right]=E\left[Y_{i t}(0)-Y_{i t-1}(0) \mid H_{i t}=1\right] \text { for all } t \geq g
\end{equation}
where we define $H_{i t}$ to be an indicator variable that takes the value 1 if the driver has not yet been treated at time $t$ (and includes never-treated drivers), and takes the value 0 otherwise.

Assumption 1 is plausible because for drivers who remain unaware of the policy that constitutes treatment, their potential outcomes and behavior remain unaffected prior to implementation. For drivers who become aware of the policy after the announcement, it is unlikely that they would significantly change their behavior before the policy took effect (i.e., receive the treatment and enroll in the program). One might argue that there are drivers that change their driving behavior before receiving treatments in anticipation of the upcoming policy change. We also test such hypotheses by setting different anticipation periods and find that the results remain consistent.

Assumption 2 requires that the average change in outcome measure of the potential untreated drivers should evolve in the same way as the never-treated drivers. Each (groupltime)-specific treatment effect is estimated by computing a weighted estimate for which the reference time period is $g-1$. Based on this specification, we can obtain the overall treatment effect by averaging all the identified cohort-time average treatment effects $\operatorname{ATT}(g, t)$ for each of the 13 treated driver cohorts at up to 25 weeks. This aggregation weights each $\operatorname{ATT}(g, t)$ by the share of treated drivers in cohort g observed at week t for $t \geq g$. Specifically, we define the overall treatment effect $\psi$ as follows:
\begin{equation}
    \psi=\frac{1}{k} \sum_{g t>g} \sum_g \omega_g \operatorname{ATT}(g, t),
\end{equation}

where $\omega_g$ is the weight of driver cohort $g$ which is proportional to the group size of the driver cohort g. $k$ normalizes the weights so that they sum to one.

Additionally, as recommended by an influential recent survey on modern difference-in-differences econometrics (\cite{roth2023whats}), we conducted the Goodman-Bacon decomposition (\cite{goodmanbacon2021variation}) and we confirm the validity of applying CSA in our setting. We provide additional details about the Goodman-Bacon decomposition results in Appendix D.  

Next, we compute dynamic treatment effects by length of exposure to the treatment. That is, we aggregate $\operatorname{ATT}(g, t)$ for the treated driver cohort g at the $e$ th week after they first enter the treatment group (i.e., $e=t-g$). Specifically, we define the average treatment effect by length of exposure parameter $e$ as follows:
\begin{equation}
    \theta(e)=\frac{1}{k} \sum_g \omega_g \operatorname{ATT}(g, g+e)
\end{equation}

where $\omega_g$ is proportional to the group size of driver cohort g , and $k$ ensures the weights sum up to one. For $e<0, \theta(e)$ captures pretreatment trend and conversely, for $e \geq 0, \theta(e)$ captures the time variation of the treatment effect.

\begin{table}[htbp]
\centering
\small
\caption{Baseline results on the rest of outcome measures}
\label{tab:c4}
\begin{tabular}{llcccc}
\toprule
\textbf{Category} & \textbf{Dependent Variable} & \textbf{Model 1} & \textbf{Model 2} & \textbf{Model 3} & \textbf{Model 4} \\
\midrule
\multirow{2}{*}{Quantity} 
    & log(num\_miles) & 0.6703*** (0.0123) & 0.4071*** (0.0361) & 0.5320*** (0.0694) & 0.1757*** (0.0482) \\
    & log(num\_sessions) & 0.2092*** (0.0043) & 0.1228*** (0.0138) & 0.1790*** (0.0258) & 0.0849*** (0.0185) \\
\midrule
\multirow{3}{*}{Utilization} 
    & log(trip\_duration) & 0.2554*** (0.0049) & 0.1621*** (0.0151) & 0.1757*** (0.0302) & 0.0357* (0.0152) \\
    & log(ave\_dur\_per\_session) & 0.1610*** (0.0029) & 0.0880*** (0.0086) & 0.1071*** (0.0158) & 0.0289* (0.0117) \\
    & log(ave\_n\_trip\_per\_hour) & 0.1842*** (0.0032) & 0.0973*** (0.0097) & 0.1850*** (0.0193) & 0.0598*** (0.0126) \\
\midrule
\multirow{3}{*}{Earnings} 
    & log(earning\_per\_ride) & 0.4445*** (0.0069) & 0.2398*** (0.0200) & 0.3845*** (0.0419) & 0.1213*** (0.0275) \\
    & log(tips) & 0.3305*** (0.0080) & 0.2082*** (0.0247) & 0.1977*** (0.0489) & 0.1091** (0.0331) \\
    & perc\_tips & 0.0056*** (0.0002) & 0.0035*** (0.0008) & 0.0031*** (0.0014) & -0.0021* (0.0009) \\
\midrule
\multirow{3}{*}{Service} 
    & log(actual\_wait\_time) & 0.0221*** (0.0003) & 0.0113*** (0.0012) & 0.0171*** (0.0021) & -0.0013 (0.0016) \\
    & weekly\_cancel\_rate & 0.0061*** (0.0004) & 0.0042*** (0.0013) & 0.0016*** (0.0028) & -0.0140 (0.0017) \\
    & driver\_rating & 0.0005*** (0.0000) & 0.0005*** (0.0000) & 0.0006*** (0.0000) & 0.0003 (0.0007) \\
\midrule
\multicolumn{2}{l}{Driver FE} & Yes & Yes & Yes & Yes \\
\multicolumn{2}{l}{Week FE} & Yes & Yes & Yes & Yes \\
\midrule
\multicolumn{2}{l}{Drivers} & 55219 & 9040 & 5055 & 1335 \\
\multicolumn{2}{l}{Observations} & 1380475 & 226000 & 252750 & 66750 \\
\bottomrule
\end{tabular}
\vspace{1em}
\begin{flushleft}
\footnotesize
\textit{Note:} Standard errors in parentheses. * p$<$0.1, ** p$<$0.05, *** p$<$0.01.
\end{flushleft}
\end{table}

\subsection{Appendix C: Constructing Matching Sample}

\begin{figure}[H]
    \centering
    \includegraphics[width=0.7\linewidth]{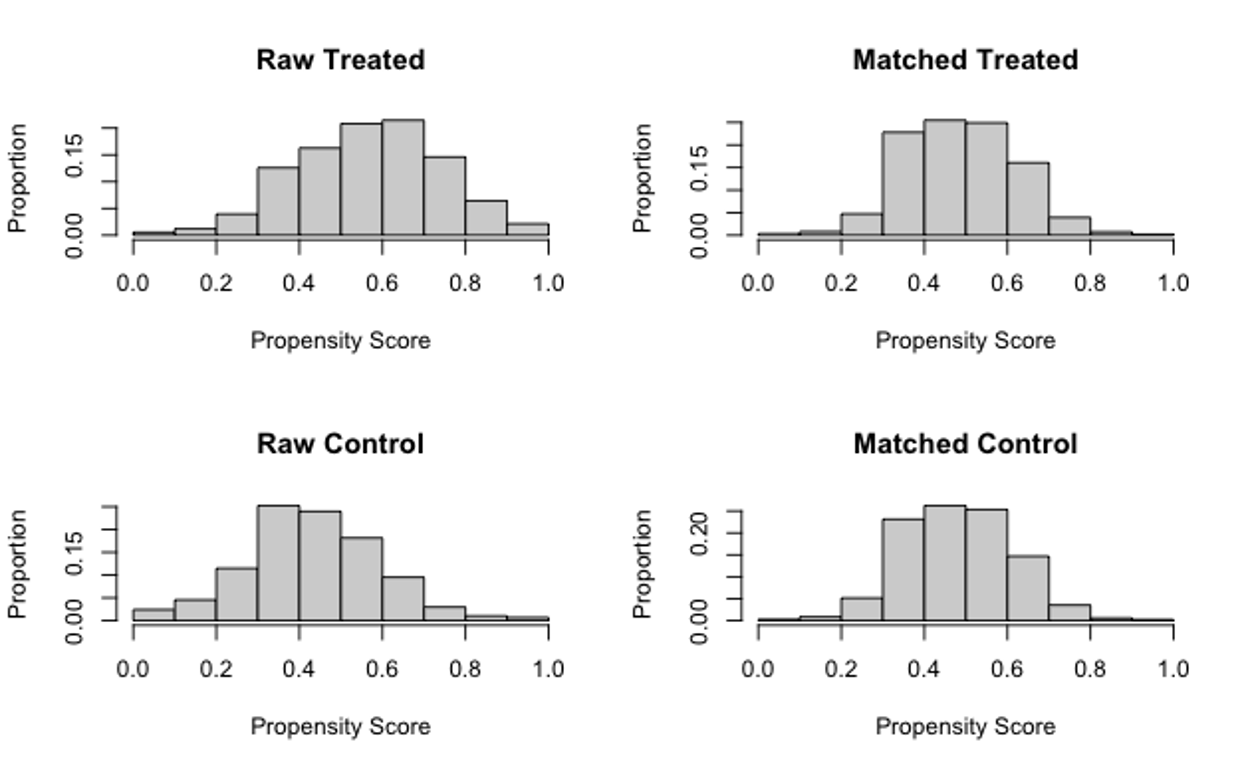}
    \caption{\centering  Distribution of Propensity Score for Treated and Control Drivers (prior year behavior) Before Matching and After Matching}
    \label{fig:balance_prior}
    \vspace{-2em}
\end{figure}

\begin{figure}[H]
    \centering
    \includegraphics[width=0.8\linewidth]{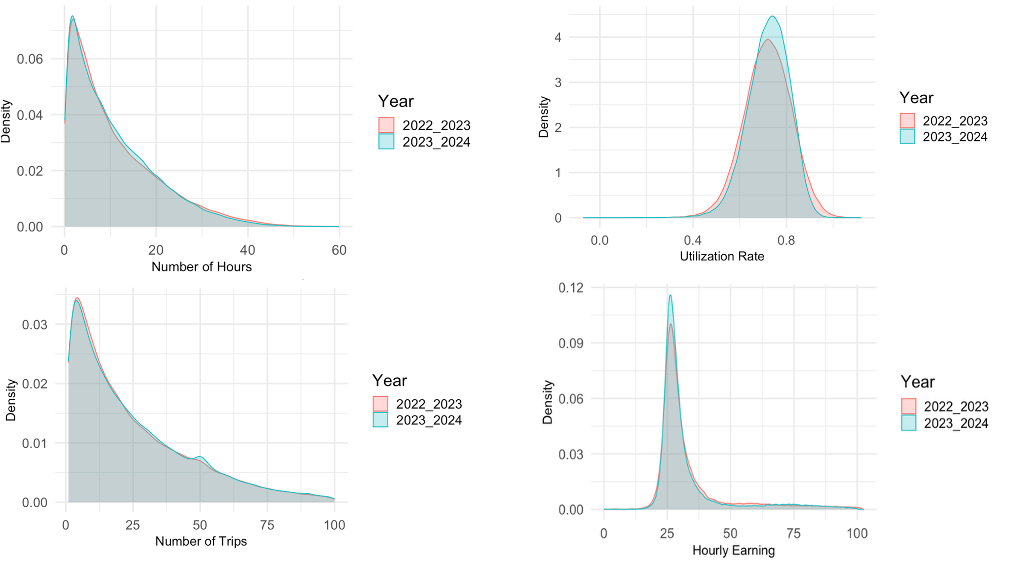}
    \caption{\centering  Distribution of Covariates After Matching (Using prior year as control)
}
    \vspace{-2em}
\end{figure}

\subsection{Appendix C: Robustness Check}
\subsubsection{Placebo test – using prior year drivers and conduct same baseline analysis}

As a placebo test, we replicate our baseline analysis using the same set of drivers but with their behavior in the prior year (2022–2023). Specifically, we retain the treatment assignment labels from the policy rollout year (2023–2024), but instead apply them to the prior-year data. In other words, drivers are classified as “treated” or “untreated” according to their eventual exposure in 2023–2024, but the outcome variables are drawn entirely from the corresponding weeks in 2022–2023. This placebo test provides a useful benchmark: if our results were being driven by pre-existing differences in trends between treated and untreated drivers, we would expect to find similar treatment effects in the prior year. By contrast, the absence of significant placebo effects strengthens the credibility of our identification strategy by showing that the estimated policy effects are not simply artifacts of underlying differential trends. We report the results in Table \ref{placebo1}.

\begin{table}[htbp]\centering
\small
\renewcommand{\arraystretch}{1.3}\begin{tabular}{lcccc}
\toprule
\textbf{Dependent Variable} & \textbf{log(num\_hours)} & \textbf{log(num\_trips)} & \textbf{Utilization rate} & \textbf{log(hourly\_earnings)} \\
\midrule
Overall treatment effect & 0.03363 & 0.03154 & 0.0044 & 0.02792 \\
& (0.0751) & (0.01155) & (0.0267) & (0.0128) \\
\midrule
Driver FE & Yes & Yes & Yes & Yes \\
Week FE & Yes & Yes & Yes & Yes \\
\midrule
Drivers & 37266 & 37266 & 37266 & 37266 \\
Observations & 931650 & 931650 & 931650 & 931650 \\
\bottomrule
\end{tabular}
\caption{Illustrates overall treatment effect using prior year drivers and conduct same baseline analysis}
\label{placebo1}
\end{table}

\subsubsection{Placebo test – SBD/OCX drivers using prior year as control}
We also conduct a placebo test using drivers who completed all pretreatment trips in the SBD and OCX markets. To ensure comparability, we restrict attention to drivers whose pretreatment behavior is stable across both the current year (2023–2024) and the prior year (2022–2023). Specifically, we require that their activity levels in the pretreatment period are similar across the two years. This yields 487 drivers.
For each driver, we then compare outcomes in their prior year with those in their current year, aligning both around the policy rollout week (week 0). Drivers are classified using the same treatment timing labels as in the baseline specification, but the “treated” outcomes now correspond to their current-year behavior. This design provides a within-driver placebo check: if the estimated treatment effects in our main results were driven by spurious year-to-year fluctuations or time-varying shocks specific to SBD/OCX markets, we would expect to observe significant effects here as well. The absence of such effects further supports the validity of our identification strategy. We report the results in Table \ref{placebo2}.

\begin{table}[htbp]\centering
\small
\renewcommand{\arraystretch}{1.3}
\begin{tabular}{lcccc}
\toprule
\textbf{Dependent Variable} & \textbf{log(num\_hours)} & \textbf{log(num\_trips)} & \textbf{Utilization rate} & \textbf{log(hourly\_earnings)} \\
\midrule
Treated\_Year × After & -0.0198 & -0.0067 & 0.0210 & 0.0766 \\
                      & (0.0364) & (0.0485) & (0.0107) & (0.0555) \\
\midrule
Driver FE             & Yes & Yes & Yes & Yes \\
Week FE               & Yes & Yes & Yes & Yes \\
\midrule
Drivers               & 487 & 487 & 487 & 487 \\
Observations          & 24350 & 24350 & 24350 & 24350 \\
\bottomrule
\end{tabular}
\caption{Illustrates overall treatment effect using drivers who drive exclusively in SBD/OCX markets in all pretreatment weeks.}
\label{placebo2}
\end{table}

\subsection{Appendix D: Bacon Decomposition}
\begin{figure}[H]
    \centering
    \includegraphics[width=0.9\linewidth]{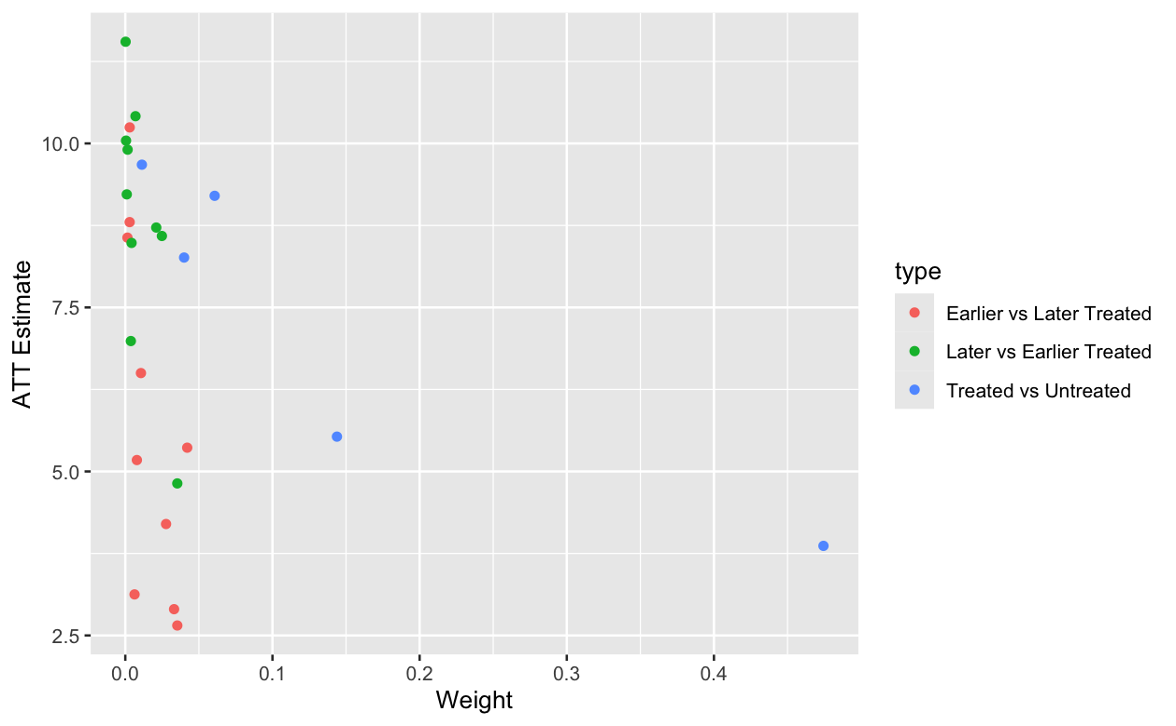}
    \caption{\centering  Bacon decomposition
}
    \label{fig:bacon}
    \vspace{-2em}
\end{figure}

The TWFE model is considered to be problematic in the staggered setting due to the heterogeneous treatment timing and driver behavior. This is because TWFE model calculates an estimate of treatment effects as a weighted average of changing combination of different treatment cohorts and its corresponding control group, which contains problematic comparison of newly treated units relative to already treated units, as it is not the path of untreated potential outcomes and instead includes treatment effect dynamics. Thus these dynamics appear in the TWFE parameter, making it very hard to give a clear causal interpretation and not robust to treatment effect heterogeneity (Borusyak and Jaravel (2018), de Chaisemartin and D’Haultfoeuille (2020), Gardner (2021), and Sun and Abraham (2021)). To show this, we conduct a Bacon Decomposition (Goodman-Bacon (2021)) that first calculates the effect derived from each of the 2x2 comparisons and its associated weights and we further aggregate the results for each of the 2x2 comparisons into three groups: (1) Earlier vs Later Treated (2) Later vs Earlier Treated and (3) Treated vs Untreated. Each estimate of these 2x2 comparisons from these three groups with its associated weights are demonstrated in Figure \ref{fig:bacon} (the fraction of these problematic comparisons between newly treated units and already treated units with their associated weights is labeled green). Table \ref{tab:bacon} highlights the aggregated measure of the treatment effect estimates from each group and its associated weights. We see that the estimated effect from (3) Treated vs Untreated is 0.493, and it accounts for a large portion (73\%) of the overall estimated effect. However, the estimated treatment effect for (2) Later vs Earlier Treated is 1.58, almost triple than the effect of group 3. Although group 2 only accounts for a relatively small size (9\%) of the overall treatment effect, the larger magnitude of the effect is likely to bias the estimation.

\begin{table}[H]
\centering
\caption{2x2 Comparison Groups: Coefficient Estimates and Weights}
\begin{tabular}{lcc}
\toprule
\textbf{2x2 Comparison Groups} & \textbf{Coefficient Estimate} & \textbf{Weight} \\
\midrule
Earlier vs Later Treated & 0.69251 & 0.17626 \\
Later vs Earlier Treated & 1.58718 & 0.09160 \\
Treated vs Untreated     & 0.49347 & 0.73214 \\
\bottomrule
\label{tab:bacon}
\end{tabular}
\end{table}

\subsection{Appendix F: K-MEANS Clustering}
We perform a k-means clustering procedure to group drivers based on a set of covariates calculated from their pretreatment period. We primarily leverage two sources of behavioral information: drivers' work windows and vehicle types. 

We divide a 24-hour day into eight work-window categories that reflect typical Lyft driver schedules---including peak demand periods, rest needs, and common behavioral patterns: 
Morning Rush (6am--9am), 
Midday Break (9am--12pm), 
Lunch/Errands (12pm--2pm), 
Afternoon Lull (2pm--4pm), 
Evening Rush (4pm--7pm), 
Dinner and Night Shift (7pm--12am), 
Late Night/Early Morning (12am--3am), and 
Rest Window (3am--6am). 

For each driver, we compute the fraction of trips occurring in each of these eight windows using their pretreatment-period data (\texttt{perc\_morning\_rush}, \texttt{perc\_midday\_break}, \texttt{perc\_lunch\_break}, \texttt{perc\_aft\_lull}, \texttt{perc\_eve\_rush}, \texttt{perc\_dinner\_shift}, \texttt{perc\_ely\_morning}, \texttt{perc\_rest\_window}). We also compute the fraction of trips happening on weekends (\texttt{perc\_wkd}). 

Drivers also operate across multiple vehicle types, providing different levels of service (standard ride, plus, premium, luxury, luxury SUV, and courier). Accordingly, we compute the fraction of rides from each vehicle type for each driver during the pretreatment period (\texttt{perc\_courier}, \texttt{perc\_lux}, \texttt{perc\_luxsuv}, \texttt{perc\_premium}, \texttt{perc\_standard}, \texttt{perc\_plus}).

\begin{figure}[H]
    \vspace{-2em}
    \centering
    \includegraphics[width=0.9\linewidth]{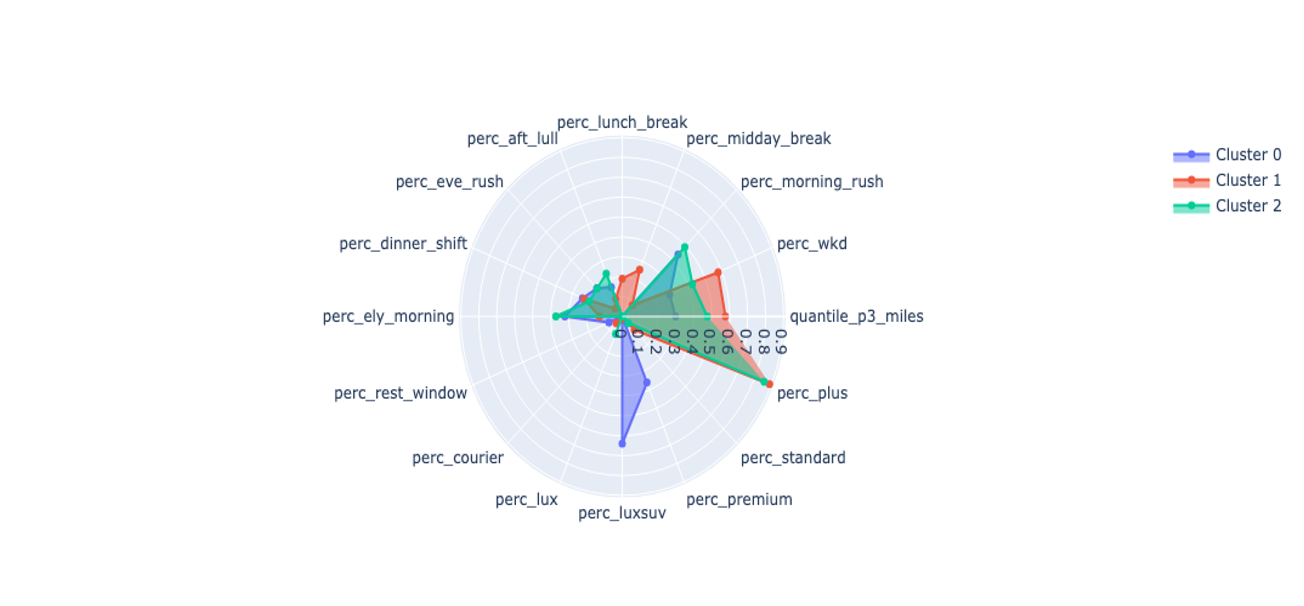}
    \caption{\centering  Driver Segmentation – Three Driver Clusters
}
    \label{fig:kmeans}
    \vspace{-2em}
\end{figure}

\subsection{Appendix G: Demand-Side Statistics }
\begin{figure}[H]
    \centering
    \includegraphics[width=0.7\linewidth]{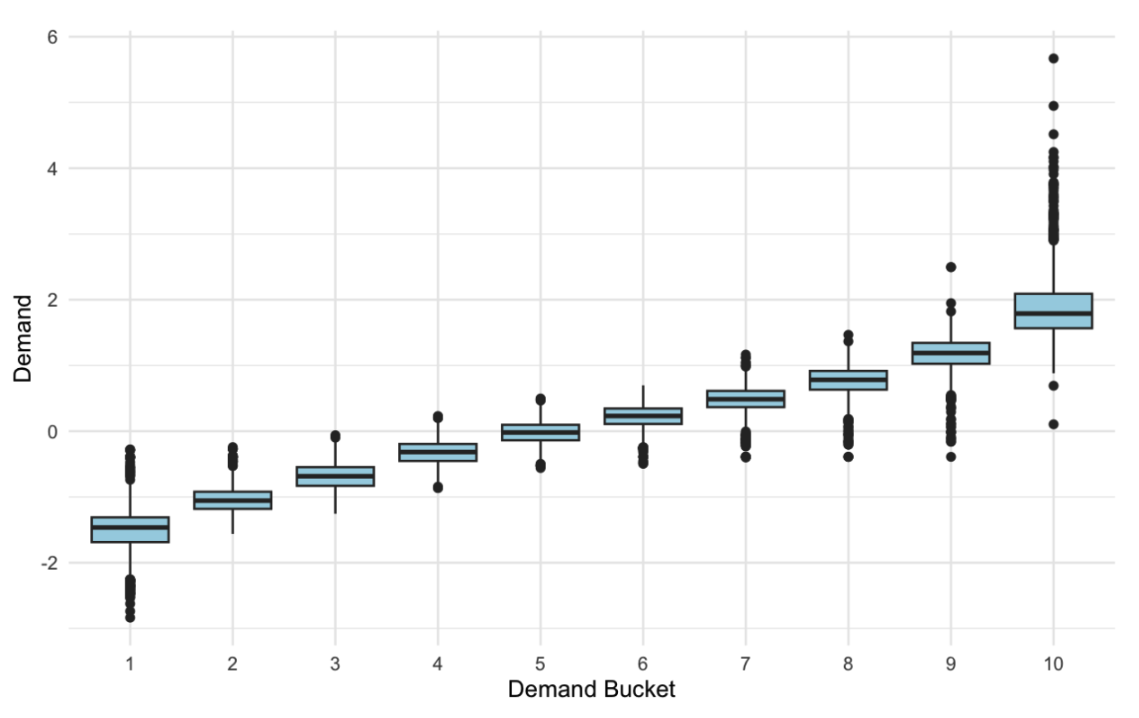}
    \caption{\centering  Distribution of Demand Intents Across Region Deciles
}
    \label{fig:demand}
    \vspace{-2em}
\end{figure}

In this section, we begin by documenting that the distribution of demand intents is highly skewed. Figure \ref{fig:demand} presents the weekly zip code–level demand intents using boxplot, normalized using z-scores to protect confidential information. The x-axis represents deciles of regions, grouped by their average weekly demand intents over 25 weeks in 2024. The y-axis shows the normalized (z-score) weekly demand intents at the zip code level. The distribution is right-skewed, indicating that a small fraction of regions account for a large share of total demand.

\begin{figure}[H]
    \centering
    \includegraphics[width=0.7\linewidth]{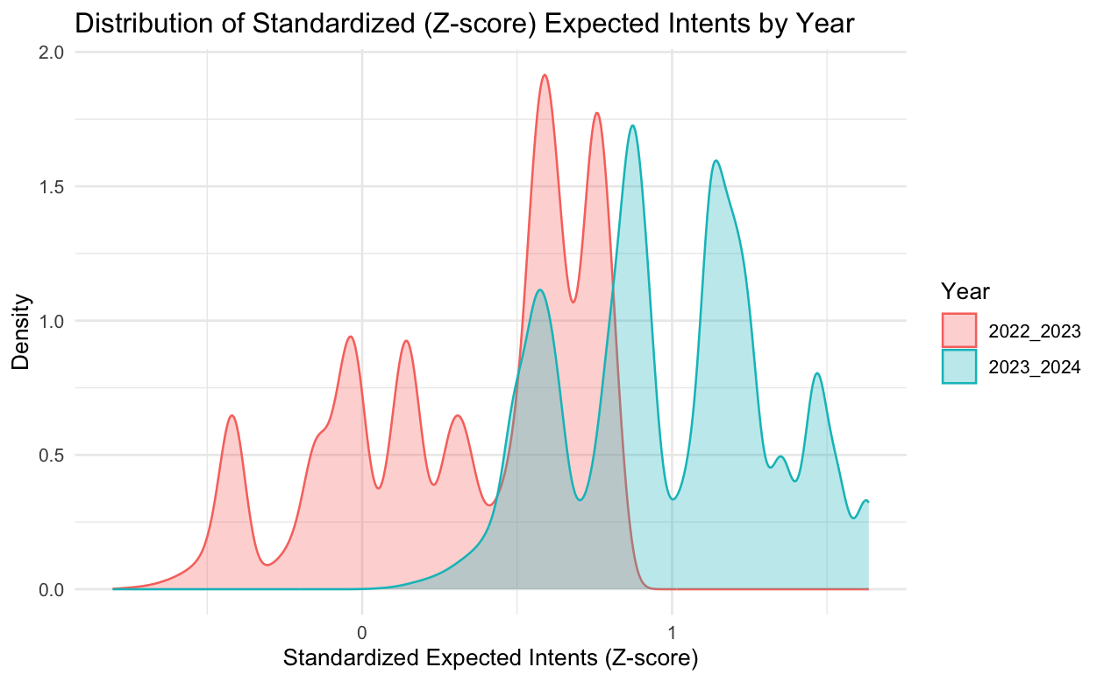}
    \caption{\centering  Distribution of expected demand intents (z-score) across two years}
    \label{fig:expected_intents}
    \vspace{-2em}
\end{figure}

Then we describe how we addressed the endogeneity issue of demand in model 4.   Since demand may be endogenous, we first calculate, for each driver, the fraction of geographic areas visited during the pre-treatment period across the two years. (as indicated in Figure \ref{fig:expected_intents}). We then compute a weighted average of demand intents for week $t$ in year $y$, using these exposure fractions as weights. This approach allows us to capture time-varying market conditions that reflect each driver’s baseline demand environment while mitigating simultaneity concerns. The impact of the treatment on driver behavior is captured by $\beta_3$, which is reported in Table 5 (column 4). 

\subsection{Appendix H: Counterfactual Analysis }

\begin{figure}[H]
    \centering
    \includegraphics[width=0.7\linewidth]{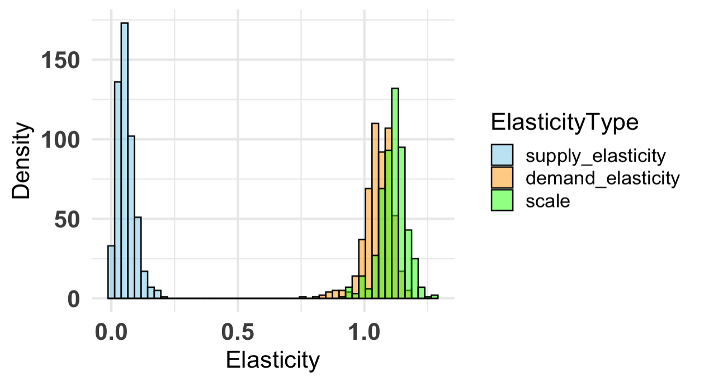}
    \caption{\centering  Distribution of Elasticities 
}
    \label{fig:Elasticities}
    \vspace{-2em}
\end{figure}

\begin{figure}[H]
    \centering
    \includegraphics[width=0.7\linewidth]{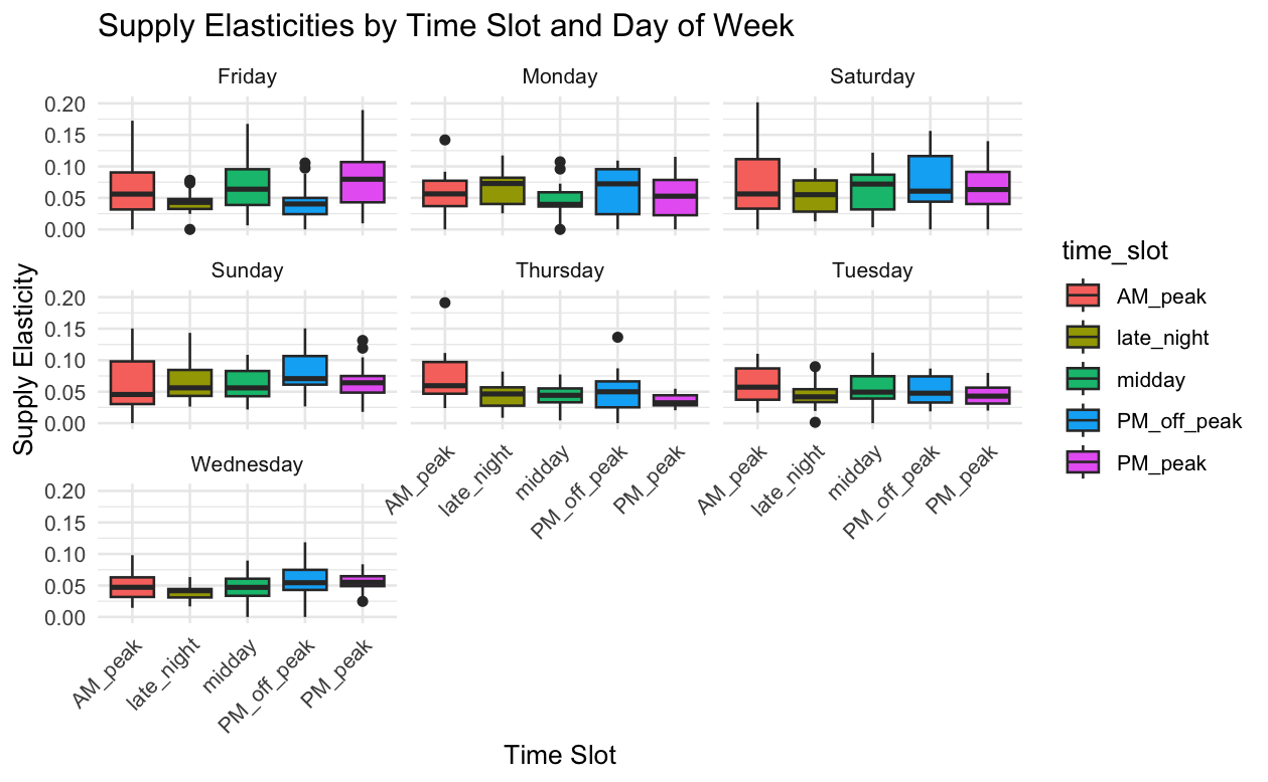}
    \caption{\centering Supply Elasticities by Time Slots and Day of Week 
}
    \label{fig:Elasticities_time}
    \vspace{-2em}
\end{figure}

\end{document}